\documentclass[a4paper]{aiaa-tc}
\AIAAsubmitinfo{Pioneer Anomaly: Nondedicated Test, Rathke and
Izzo}

\usepackage{indentfirst}
\usepackage{amsmath}
\usepackage{amssymb}

\newcommand{\nn}{\nonumber}

\def\centereps#1#2{\centerline{\includegraphics[width=#1]{#2}}}
\def\vett#1{\mathbf{#1}}

\title{\normalsize{Options for a nondedicated mission to test the Pioneer
anomaly}}

\author{\small%
Andreas Rathke%
\thanks{System engineer, EADS Astrium GmbH, Dept.~AED41
88039 Friedrichshafen, Germany. \mbox{Email: andreas.rathke@astrium.eads.net}}
\hskip.2cm and \hskip.2cm Dario Izzo%
\thanks{Research Fellow, Advanced Concepts Team, European Space Agency, ESTEC, Keplerlaan
1, 2201 AZ Noordwijk, The Netherlands. \mbox{Email: dario.izzo@esa.int}}}

\begin{document}
\maketitle

\begin{abstract}
The Doppler-tracking data of the Pioneer 10 and 11 spacecraft show an
unmodelled constant acceleration in the direction of the inner Solar
System. Serious efforts have been undertaken to find a conventional
explanation for this effect, all without success at the time of
writing. Hence the effect, commonly dubbed the Pioneer anomaly, is
attracting considerable attention. Unfortunately, no other space
mission has reached the long-term navigation accuracy to yield an
independent test of the effect. To fill this gap we discuss strategies
for an experimental verification of the anomaly via an upcoming space
mission.  Emphasis is put on two plausible scenarios:
nondedicated concepts employing either a planetary exploration
mission to the outer Solar System or a piggybacked micro-spacecraft to
be launched from a mother spacecraft travelling to Saturn or
Jupiter. The study analyses the impact of a Pioneer anomaly test on
the system and trajectory design for these two paradigms. It is found
that both paradigms are capable of verifying the Pioneer anomaly and
determine its magnitude at 10\% level. Moreover the concepts can
discriminate between the most plausible classes of models of the anomaly,
a central force, a blueshift of the radio signal and a drag-like
force.  The necessary adaptions of the system and mission design
do not impair the planetary exploration goals of the missions.
\end{abstract}

\section*{Nomenclature}
\begin{tabbing}
\hspace*{1.5cm}\= \hspace*{1.5cm}\=\kill
$A_{S/C}$ \> = \> crossectional area of the spacecraft, m$^2$\\
$a_\odot$ \> = \> acceleration due to solar radiation pressure, m/s$^2$\\
$a_H$ \> = \> Hubble acceleration, m/s$^2$\\
$c$ \> = \> speed of light, $\approx 3 \times 10^8\,{\rm m/s}$\\
$\vec{e}_A$ \> = \> unit vector normal to the area $A$\\
$\vec{e}_\odot$ \> = \> unit vector pointing towards Sun\\
$F$ \> = \> force, N\\
$f$ \> = \> a generic tracking observable\\
$g_0$ \> = \> gravitational acceleration at the Earth's surface, m/s$^2$\\
$I$ \> = \> moment of inertia, kg\,m$^2$ \\
$I_{sp}$ \> = \> specific impulse, s\\
$k$ \> = \> Boltzmann constant, $1.3806 \times 10^{-23}$\,J/K\\
$M_\alpha$ \> = \> total mass of $\alpha$-particles produced by
radioactive
decay, kg\\
$M_{S/C}$ \> = \> spacecraft wet mass, kg\\
$H_0$ \> = \> Hubble constant, km/s/Mpc\\
$P_a$ \> = \> asymmetrically radiated power, W\\
$P_\text{tot}$ \> = \> total radiated power, W\\
$P_\odot$ \> = \> solar radiation constant $1367\,{\rm W\,AU^2/m^2}$\\
$R$ \> = \> individual gas constant, J/kgK\\
$r$ \> = \> heliocentric distance, m  \\
$r_\oplus$ \> = \> mean radius of Earth orbit, m\\
$r_{p}$ \> = \> radius of pericentre from planet $P$, km \\
$s$ \> = \> geocentric distance of spacecraft, km\\
$s^*$ \> = \> deviation from nominal spacecraft trajectory, km\\
$T$ \> = \> temperature, K\\
$T_\text{tank}$ \> = \> temperature of fuel in tank, K\\
$T_s$ \> = \>  stagnation temperature, K\\
$T_0$ \> = \> temperature at nominal emissivity, K\\
$t$ \> = \> time, s\\
$t_e$ \> = \> time of departure at Earth, MJD\\
$t_p$ \> = \> time of arrival/swingby at planet, MJD\\
$t_\oplus$ \> = \> orbital period of the Earth, s\\
$V_P$ \> = \> heliocentric velocity of planet $P$, km/s\\
$v_\text{in}$ \> = \> inbound asymptotic velocity, km/s\\
$v_\text{out}$  \> = \> outbound asymptotic velocity, km/s\\
$v_\alpha$ \> = \> velocity of $\alpha$-particles, m/s\\
$v_\oplus$ \> = \> mean heliocentric velocity of the Earth, km/s\\
$\alpha_\oplus$ \> = \> longitude in geocentric ecliptic coordinate system, deg\\
$\alpha^*$ \> = \> deviation from nominal geocentric azimuth angle, deg\\
$\beta$ \> = \> angle between Earth--spacecraft direction and direction of anomaly, deg\\
$\beta_\odot$ \> = \> angle between Sun--spacecraft direction and direction of anomaly, deg\\
$\beta_\oplus$ \> = \> Earth--spacecraft--Sun angle, deg\\
$\gamma$ \> = \> flight angle (angle between velocity vector and local horizontal), deg\\
$\Delta a$ \> = \> systematic uncertainty of acceleration $a$, m/s$^2$\\
$\Delta f$ \> = \> uncertainty on the generic tracking observable $f$\\
$\Delta s$ \> = \> systematic uncertainty on the geocentric distance $s$, km\\
$\Delta v$ \> = \> systematic uncertainty on the velocity $v$, km/s\\
$\Delta M$ \> = \> mass of expelled propellant, kg\\
$\Delta V$ \> = \> velocity increment, km/s\\
$\Delta \epsilon$ \> = \>change of emissivity\\
$\Delta \epsilon_{max}$ \> = \>maximal change of emissivity\\
$\Delta \mu_\odot$ \> = \>change of the effective reduced Solar mass, km$^3$/s$^2$\\
$\epsilon_0$ \> = \>nominal emissivity\\
$\eta$ \> = \>specular reflectivity\\
$\theta$ \> = \>angle enclosed by $\vec{e}_A$ and $\vec{e}_\odot$, deg\\
$\kappa$ \> = \>adiabatic exponent\\
$\mu_P$ \> = \>reduced mass of planet $P$, km$^3$/s$^2$\\
$\mu_\odot$ \> = \> reduced solar mass, km$^3$/s$^2$\\
$\rho$ \> = \>true heliocentric distance, km\\
$\psi$ \> = \>azimuth angle of cylinder coordinates, deg\\
$\phi$ \> = \>mean anomaly of Earth orbit, deg\\
$\sigma$ \> = \>standard deviation\\
$\omega$ \> = \>rotational velocity of spacecraft, rad/s\\
\end{tabbing}
{\em Superscripts}
\begin{tabbing}
\hspace*{1.5cm}\= \hspace*{1.5cm}\=\kill
$^*$ \> = \>anomalous\\
\end{tabbing}
{\em Subscripts}
\begin{tabbing}
\hspace*{1.5cm}\= \hspace*{1.5cm}\=\kill
$_\text{track}$ \> = \>tracking error\\
$_0$ \> = \>at time $t=0$, i.\,e.\ beginning of measurement\\
$_{\|}$ \> = \>parallel to Earth--spacecraft vector\\
$_{\bot}$ \> = \>orthogonal to Earth--spacecraft vector\\
$_\odot$ \> = \> solar\\
$_\oplus$ \> = \> Earth/terrestrial
\end{tabbing}

\section{Introduction}

Doppler tracking data of the Pioneer 10 and 11 deep-space probes show
a deviation between the orbit reconstruction of the spacecraft and
their Doppler tracking signals
\cite{Anderson:1998jd,Anderson:2001sg}. This discrepancy, that has
become known as the Pioneer anomaly, can correspond either to a small
constant deceleration of the spacecraft of roughly $9\times
10^{-10}\,{\rm m/s^2}$ or to an anomalous blueshift of the radio
signal at a rate of $6 \times 10^{-9}\,{\rm Hz/s}$.  Since no
unambiguous conventional mechanism to explain the anomaly, such as an
on-board force, has been identified there is a growing number of
studies, which consider an explanation in terms of a novel physical
effect.

In April 2004 the European Space Agency (ESA) invited the scientific
community to participate in a Call for Themes for Cosmic Vision
2015-2025, to assist in developing the future plans of the Cosmic
Vision programme of the ESA Directorate of Science. Among the 32
proposals received in the field of Fundamental Physics, five were
proposing a space experiment to investigate the Pioneer anomaly. In
its recommendation for the Cosmic Vision programme, the Fundamental
Physics Advisory Group (FPAG) of ESA considered these proposals as
interesting for further investigation.\cite{FPAG} In view of the
controversial discussion still surrounding the effect on the one hand
and its high potential relevance for our understanding of the laws of
physics on the other hand, the FPAG recommended that ESA should study the
possibility to investigate the putative anomaly on board of a
nondedicated exploration mission.

Motivated by this important discussion we provide a preliminary
assessment of the capabilities of missions to the outer Solar System
to investigate the Pioneer anomaly. We identify two classes of
mission that could well represent a future exploration mission. The
first class is that of low-mass low-thrust orbiter missions to the
outer planets. The second class is that of a heavy, nuclear-reactor
powered spacecraft, as formerly proposed by NASA's Prometheus Programme, to
explore the giant planets. Within these two paradigms we analyse
missions to all planets from Jupiter outward and consider to what
extent a verification and characterisation of the Pioneer anomaly is
possible.

The layout of our considerations is the following: We begin with a
review of the Pioneer anomaly in Sec.~\ref{PA}.  After a description
of the observed anomaly in the Pioneer tracking in
Sec.~\ref{PA}.\ref{PAa} we turn to the considerations that have been
put forward to explain the anomaly in terms of systematics in
Sec.~\ref{PA}.\ref{PAb}. In Sec.~\ref{PA}.\ref{PAc} we review
approaches to explain the anomaly as a
novel physical effect. This review leads us to the formulation of the
experimental requirements, that a mission to test the Pioneer anomaly,
has to fulfil, in Sec.~\ref{PA}.\ref{PAe}.  In Sec.~\ref{PA}.\ref{PAd}
we discuss the navigational accuracy of past and present
deep-space missions and explain why none of these mission is likely to
decide the issue if the Pioneer anomaly is indeed of physical
significance. Sec.~\ref{nonded} turns to the discussion of
nondedicated mission concepts for a test of the Pioneer anomaly. We
start by discussing the major design drivers for missions to the outer
Solar System in Sec.~\ref{nonded}.\ref{capa}. Then
Sec.~\ref{nonded}.\ref{pop} and \ref{nonded}.\ref{pig} give an
overview of the two scenarios that we consider.  Sec.~\ref{des}
discusses in detail the necessary design considerations to reduce the
systematic accelerations onboard a deep-space probe to a tolerable
amount for a test of the Pioneer anomaly. In particular the aspects of
thrust history uncertainties (Sec.~\ref{des}.\ref{desa}), fuel leaks
and outgassing (Sec.~\ref{des}.\ref{desb}), thermal radiation
(Sec. \ref{des}.\ref{desc}), the radio-beam reaction force
(Sec.~\ref{des}.\ref{desd}) and solar radiation pressure
(Sec.~\ref{des}.\ref{dese}) are addressed. Sec.~\ref{des}.\ref{err}
summaries the estimated error budget and Sec.~\ref{des}.\ref{dessum}
summarises the necessary modifications in the spacecraft design to
fulfil the test requirements.  The second major topic is the
development of a measurement strategy for the test in
Sec.~\ref{trac}. Section \ref{trac}.\ref{inst} investigates the
instrumentation options for an verification of the anomaly. It is
found that the experiment will have to rely on radio tracking. Consequently we
 review the available radio tracking methods in
Sec.~\ref{trac}.\ref{meth}. This is followed by a discussion of the
relevant tracking observable in Sec.~\ref{trac}.\ref{obs}. In
Sec.~\ref{trac}.\ref{perf} the radio-tracking performance of the two
mission paradigms is estimated. Based on the design and mission
requirements obtained, the space of trajectory options is explored in
Sec.~\ref{traj}. This is done separately for the two mission paradigms
in Sec.~\ref{traj}.\ref{trajpop} and \ref{traj}.\ref{trajpig}. The
conclusions of our analysis are summarized in Sec.~\ref{conc}.

\section{The Pioneer anomaly \label{PA}}

\subsection{The tracking-data anomaly \label{PAa}}

The Pioneer~10 and 11 spacecraft, launched on 2 March 1972 and 5 April
1973, respectively, were the first to explore the outer Solar System
(see Lasher and Dyer\cite{Pioneers} for an overview of the Pioneer 10 and 11
missions.). Since its Jupiter gravity assist on 4 December 1973
Pioneer~10 is on a hyperbolic coast.  In the heliocentric J2000
reference frame the ascending node of the asymptote was (and has since
remained) $-3.4\,$deg; the inclination of the orbit is $26.2$\,deg.
Pioneer~11 used a Saturn swingby on 1 September 1979 to reach a
hyperbola with an asymptotic ascending node of $35.6$\,deg and an
inclination of $9.5$\,deg. The orbit determination for both craft
relied entirely on Doppler tracking.

Already before the Jupiter
swingby, the orbit reconstruction for Pioneer 10 indicated an
unmodelled deceleration of the order of $10^{-9}\,{\rm m/s^2}$ as
first reported by Null.\cite{Null} This effect was, at that time,
attributed to on-board generated systematics (i.e. unmodelled
behaviours of the spacecraft systems), in particular to fuel
leaks. However an unmodelled deceleration remained also during the
hyperbolic coast, although the number of attitude-control manoeuvres
was reduced to approximately one every five months. Hence fuel
leakage, triggered by thruster activity, could no longer be considered
as an explanation. Even more surprising, the Doppler tracking of
Pioneer 11 also shows an unmodelled deceleration of a similar
magnitude. 

The anomaly on both probes has been subject to three
independent analyses that used different orbit determination
programs.\cite{Anderson:1998jd, Anderson:2001sg,Markwardt} The
conclusion of all these investigations was that an anomalous Doppler
blueshift is present in the tracking data of both craft, and that the
magnitude of the blueshift is approximately $1.1 \times 10^{-8}$\,
Hz/s, corresponding to an apparent deceleration of the spacecraft of
approximately $9 \times 10^{-10}\,{\rm m/s^2}$.  It is worth
emphasising that from the Doppler data alone it is not possible to
distinguish between an anomalous frequency shift of the radio signal
(in conventional terms this could also indicate a drift of the Deep
Space Network clocks) and a real deceleration of the spacecraft
(cf.\ Sec.~\ref{trac}.\ref{obs} below).  The observational data and the
subsequent analysis are described in detail in the work of Anderson et
al.\cite{Anderson:2001sg} and Markwardt.\cite{Markwardt} The results
of these different analyses show a discrepancy at a level of
approximately 5\% of the inferred deceleration. Unfortunately, none of
the analyses performed made use of the entire data set available.

The quality of the data is best judged from the plot of the Pioneer 10
anomalous acceleration as determined by the CHASMP software (developed
by the Aerospace Corporation) and reported by
Anderson et al.\cite{Anderson:2001sg} Fig.~9. While it is quite obvious that the
data show the existence of an anomalous acceleration it is also
obvious that the variation of the measured anomaly due to systematics
is too big to evaluate the first derivative of the anomaly. This noise
is reflected in the large overall error for the value of the anomaly
given by Anderson et al.\cite{Anderson:2001sg} $\Delta a^* = 1.33
\times 10^{-10}\,{\rm m/s^2}$. Nevertheless the deviation from the
nominal Doppler shift is highly significant: The orbit reconstruction
of Pioneer~10 is incompatible with the nominal orbit at
6$\sigma$ level.\cite{Nieto:2003rq}

\subsection{Systematics? \label{PAb}}

Many attempts\cite{Katz:1998ew,Murphy:1998hp,Scheffer1
,Scheffer2,Scheffer3,Anderson:2001ks,Anderson:2003rd,Mashhoon:2002fq,Hauck:2003gy}
have been made to interpret the anomaly as an effect of on-board
systematics ranging from fuel leakage to heat radiating from the
spacecraft. Unfortunately, the conclusions of the various studies are
far from unanimous.  In the work of Anderson et
al.\cite{Anderson:2001sg} it is concluded that none of the effects
considered is likely to have caused the anomaly. They argue that a
heat-generated anomaly would be mainly due to the heat of the
radioisotope thermoelectric generators (RTGs), and that this can be
excluded because the heat decay from the Plutonium half-life of 87.7
years, would have shown up as a decrease of the deceleration in the
longest analysed data interval for Pioneer 10, ranging from January
1987 to July 1998.

They note that gas leaks can be excluded as the cause
of the anomalous deceleration, under the sole assumption that the
amount of fuel leakage is uncorrelated between 
Pioneer 10 and 11. However, because both spacecraft designs are
identical, two identical gas leaks can ultimately not be excluded.

At the current stage of investigation it is not even
clear if one should attribute the anomaly to a conventional effect or
consider explanations rooted in new physical phenomena. A complete
examination of the full archive of Doppler data is certainly
needed. Nevertheless, even with this enhanced knowledge it seems
highly doubtful that the issue can be decided, since there exist
considerable uncertainties in the modelling of forces generated on
board Pioneer 10 and 11.  In view of the necessity for an improved
evaluation of the Doppler data, the authors feel obliged to express
their unease about the discrepancies between the results obtained with
the different orbit determination programs. In particular it is
noteworthy that the disagreement between the three analyses is bigger
that their nominal errors.

\subsection{New physics? \label{PAc}}

The inability to explain the Pioneer anomaly with conventional physics
has contributed to the growing discussion about its origin. The
possibility that it could come from a new physical effect is now being
seriously considered. In particular the coincidence in magnitude of
the Pioneer anomaly and the Hubble acceleration has stirred the
suggestion that the Pioneer anomaly could be related to the
cosmological expansion.

Although the Pioneer anomaly is an effect at the border of what is
detectable with radiometric tracking of a deep-space probe, it is huge
in physical terms. The anomaly exceeds by five orders of magnitude the
corrections to Newtonian motion predicted by general relativity (at
50\,AU solar distance).\footnote{The leading order relativistic
correction is $\sim F_{\rm N} v^2/c^2$, where $F_N$ denotes the
Newtonian gravitational force, cf.\ e.\,g.\ Montenbruck and
Gill\cite{montenbruck} p.~110f.} Hence, if the effect is not due to
systematics, it would have a considerable impact on our models of
fundamental forces, regardless of whether the anomaly was due to a
deceleration of the spacecraft or a blueshift of the radio signal. 

One of the obstacles to an explanation of the Pioneer
anomaly in terms of new physics is that a modification of gravitation,
large enough to explain the Pioneer anomaly, easily runs into
contradiction to the planetary ephemerides. This becomes particularly
clear if one considers the orbit of Neptune. At 30\,AU, the Pioneer
anomaly is visible in the Doppler data of both Pioneer 10 and 11. The
influence of an additional radial acceleration of $a^* = 9 \times
10^{-10}\,{\rm m/s^2}$ on Neptune is conveniently parameterised in a
change of the effective reduced solar mass, $\mu_{\odot}$, felt by the
planet.\cite{Talmadge:1988qz} The resulting value, $\Delta \mu_{\odot} =
a^* r^2 = 1.4 \times 10^{-4} \, \mu_{\odot}$, is nearly two orders of
magnitude beyond the current observational constraint of $\Delta
\mu_{\odot} = ( -1.9 \pm 1.8 ) \times 10^{-6} \,
\mu_{\odot}$.\cite{Anderson:1995dw} Similarly, the Pioneer 11 data
contradict the Uranus ephemerides by more than one order of
magnitude. Thus, the Pioneer anomaly can hardly be ascribed to a
gravitational force, since this would indicate a considerable violation
of the weak equivalence principle.  In particular, planetary
constraints rule out an explanation in terms of a long-range Yukawa
force.\cite{Anderson:2001sg,Reynaud:2005xz} 

Already in the first paper discussing the Pioneer anomaly it was noted
that the magnitude of the effect coincides with the Hubble
acceleration and with the so-called Modified Newtonian Dynamics (MOND)
parameter.\cite{Anderson:1998jd} Subsequently there have been several
attempts to associate the Pioneer anomaly both with the cosmic
expansion and with the MOND model.  The Hubble acceleration $a_H$ is
formed by converting the Hubble expansion rate $H_0$,\cite{PDG2004} to
an acceleration by multiplying it by the speed of light, $a_H
\equiv cH_0 = (6.9 \pm 0.7) \times 10^{-10}\,{\rm m/s^2}$.\footnote{The
Hubble acceleration is by no means an artificial construct but is
related to actual observables. For instance it describes the lowest
order correction from the cosmic expansion to the length of light-rays
from a past event to a present-day observer $d = c \, \Delta t + \frac
{a_H}2 (\Delta t)^2$.}  Attempts to connect the Pioneer anomaly with
the cosmic expansion consider both possibilities, that the Pioneer
anomaly only affects light
propagation,\cite{Rosales:1998mj,Rosales:2002ui,Rosales:2004kb,Rosales:2005,
Ranada:2003ax,Ranada:2004mf} or that it causes a real deceleration of
the spacecraft.\cite{Modanese:1999gm,Nottale:2003zj} However, the
predominant opinion, starting with the work of Einstein and
Straus,\cite{EinsteinStraus} is that cosmic dynamics has far too little
influence to be visible in any physical processes in the Solar
System. The case has recently been reviewed, confirming the common
opinion.\cite{Cooperstock:1998ny} Other problems with this approach
are the apparent violation of the weak equivalence principle
associated with the Pioneer anomaly, and the opposite signs of the
cosmic expansion and of the Pioneer anomaly.

Modified Newtonian Dynamics (MOND) is a long-distance modification of
Newtonian gravity that successfully explains the dynamics on
galactic scales without invoking dark matter\cite{Milgrom:1983ca}
(see Sanders and McGaugh\cite{Sanders:2002pf} for a review). The MOND
parameter, $(1.2 \pm 0.3)\times 10^{-10}\,{\rm m/s^2}$, gives the
acceleration scale, at which the gravitational force changes from the
Newtonian law to the MOND law, that predicts stronger gravitational
attraction. While MOND is consistent and successful as a
non-relativistic theory, its relativistic generalisations remain
unsatisfactory because they require a fixed background structure or
even have acausal features.\cite{Bekenstein:2004ne} The Pioneer
anomaly can be connected with MOND if one assumes that the
transition between the Newtonian and MOND regimes can be
approximated by a Taylor series around the Newtonian potential and
that the MOND parameter sets the magnitude of the first term in this
Taylor expansion.\cite{Bekenstein:2004ne} Similarly the flatness of
galactic rotation curves and the Pioneer anomaly could be connected
in a gravitational theory based on a non-symmetric
metric.\cite{Moffat:2004ud}

In order to circumvent the constraints from planetary ephemerides,
mo\-men\-tum-de\-pen\-dent ``nonlocal'' modifications of general
relativity have also been
considered.\cite{Milgrom:2002tu,Jaekel:2004sj,Jaekel:2005qe} Whereas
the original idea is rather vague,\cite{Milgrom:2002tu} a more
elaborate model\cite{Jaekel:2004sj,Jaekel:2005qe} faces several
problems. Jaekel and Reynaud\cite{Jaekel:2004sj,Jaekel:2005qe}
introduced two different momentum-dependent gravitational constants
for the trace and the conformal sector of the Einstein equations. Such
running couplings lead to a violation of the Bianchi identities unless
one resorts to a non-local reformulation of the Einstein-Hilbert
action.\cite{Barvinsky:2002kh,Barvinsky:2003kg} Even then causality of
the resulting physical laws needs careful consideration. Even worse,
this modification results in an unstable dipole-ghost (cf.\
Smilga\cite{Smilga:2004cy}). It seems hard to conceive that the
combination of instability and fine-tuning between the scalar and
conformal sectors can result in a viable model.

There are several other works pursuing even more unusual lines of
explanation. The reader is referred to the papers by Anderson et
al.\cite{Anderson:2001sg} and by Bertolami and
Paramos\cite{Bertolami:2003ui} for reviews of some of the proposed
explanations of the Pioneer anomaly that rely on more exotic
physics. The models considering a blueshift of the radio signal are
reviewed by Defr\`ere and Rathke\cite{Defrere:2005ef}. Up to now all
the models to explain the Pioneer anomaly in terms of new physics
still have to be considered as incomplete. In view of the current
rapid development of the field this one might however expect
considerable progress in the next few years.

\subsection{Experimental requirements for a new test\label{PAe}}

From the analysis of the Pioneer tracking data and the theoretical
approaches at their explanation one can deduce the requirements for a
new test of the anomaly. For a verification of the anomaly one would
need a spacecraft with an acceleration systematics below the magnitude
of the anomaly. A long lasting ballistic phase in the trajectory is
mandatory so that the search for the anomaly is not overwhelmed
by thruster activity. Furthermore since it is unknown if the anomaly
is generated by a force or by an anomalous blueshift of the radio
signal the experiment has to be sensitive to both possibilities.

These generic requirements may be amended by model dependent
requirements stemming from the theoretical analysis of the anomaly. If
the anomaly is caused by a modification of the gravitational laws it
would require a violation of the weak equivalence principle. The most
plausible realisation of this would be via a momentum dependence of
the gravitational attraction. To be sensitive to such an effect one
requires a high radial velocity of the spacecraft with respect to the
Sun. This corresponds to a highly eccentric, preferably hyperbolic,
trajectory of the spacecraft.

An explicit dependence of the anomalous force on the position
of the spacecraft within the Solar System is highly improbable. This
follows from the observation that the anomalies on both Pioneer
probes do not change significantly with the position of the
spacecraft along their orbits (a small change cannot be excluded due
to the large error margin of the data); and that the trajectories of
the two Pioneers are heading away from the Sun in approximately
opposite directions and at considerably different inclinations, thus
making it possible to conclude that if such a dependence exists,
then it has to be so small as to be undetectable from study of the
Pioneer data. 

One might also envisage that the spin of the spacecraft has an
influence on the magnitude of the anomalous force (see Refs.
\cite{Anderson:2003rd,Mashhoon:2002fq,Hauck:2003gy} for an
unsuccessful attempt, which tried to locate the origin of the
anomaly in the rotation of the Pioneer probes). Such a dependence
may be reasonably excluded. The rotational speed of the Pioneer~10
spacecraft was 0.075-0.070/s; that of Pioneer 11 was about 0.122-0.120/s.
Assuming a power-law dependence of the
anomalous acceleration $a$ on the rotational velocities of the
spacecraft $\omega$, $a^* = \text{const} \, \omega^x$\,, the exponent
is constrained by the error margin of the anomalous acceleration
to $|x| < 0.7$\,. Thus, in particular, a linear dependence of the
anomalous acceleration on the rotational velocity, and a linear
dependence of the anomalous acceleration on the rotational energy of
the spacecraft, $E_\text{rot} = I \omega^2/2$ with $I$ being the
moment of inertia along the spin axis, is ruled out. Hence, a
dependence of the anomaly on the rotational parameters of the
spacecraft seems rather unlikely and in the following no
requirements on the rotational velocity will be considered.

One might want to augment the above requirements for a verification of
the anomaly by requirements that would allow a further characterisation
of the anomaly. In particular, it would be of great interest to test if
the anomaly is caused by a force of gravitational type, i.\,e.\ ``new
physics'', or of non-gravitational type, ``systematics''. Of course an
improved acceleration sensitivity of the spacecraft might allow a
determination of the force-law that generates the anomaly, e.\,g.\ its
gradient. 

Before we turn to the implementation of high acceleration
sensitivity in the design an exploration spacecraft in Sec.~\ref{des} we
consider the performance of several past, present and upcoming
deep-space mission for a test of the Pioneer anomaly.

\subsection{Other spacecraft \label{PAd}}

It stands to reason that if the anomaly detected in the tracking data
of the Pioneers, ware due to some unknown fundamental physical
phenomenon, the anomaly should be observed in the data from
other missions as well. For various reasons, up to now no other mission has
reached the long-term navigational accuracy of the Pioneer 10 and 11
spacecraft. Here we identify the design characteristics that led to
the lower navigational performance of the other past missions to the
outer Solar System and discuss the performance expectations for
current missions, which have not been designed with a test of the
Pioneer anomaly as a (secondary) mission goal.

This issue has already been analysed in detail for the Voyager
spacecraft and for Galileo and Ulysses.\cite{Anderson:2001sg} The
basic conclusion is that the 3-axis stabilisation system of the
Voyager probes performs so many attitude-control manoeuvres that it is
impossible to detect the anomalous acceleration on these
spacecraft. For Galileo and Ulysses the
large systematic errors due to solar radiation pressure and 
malfunctions of part of the attitude control systems prohibited any
reliable result. 

Also the Cassini tracking does not yield results of the necessary
precision because the spacecraft is 3-axis
stabilised.\cite{Nieto:2003rq} Furthermore thermal radiation from the
RTGs causes a large acceleration bias, the magnitude of which is not
well determined. The large bias originates from the placement of
the RTG's close to the spacecraft bus. The thermal control of the propulsion
module subsystem is accomplished by collecting thermal radiation from
the RTGs in a cavity covered with insulating blankets.\cite{Mireles}
The radiation geometry of the cavity is complicated and leads to a
large uncertainty in the acceleration bias due to RTG heat.

Amongst the current missions, ESA's Rosetta mission\cite{Rosetta1} to
the comet Churymov-Gerasimenko has a trajectory to the outer Solar
System, that would seem suited for verifying the Pioneer anomaly. The
Rosetta trajectory has a long elliptical coast arc from July 2011 to
January 2014, during which the distance from the Sun will increase
from 4.5 to 5.4\,AU.  Unfortunately the system design and operations
of the spacecraft will not allow a successful test of the Pioneer
anomaly.  During the coast arc, the Rosetta craft will enter a
so-called hibernation mode, when the power generated by the solar
arrays drops below a certain value. In this mode the spacecraft will
be spin-stabilised with a rotational velocity of approximately 1\,rpm.
Most on-board instruments, including the attitude control and radio
transmission system, will be switched-off. Unfortunately, during the
hibernation no tracking can be performed, hence the presence of a
force can only be inferred from the trajectory evolution between the
entry and exit of hibernation. The large $68\,{\rm m^2}$ solar arrays
on the craft will cause an acceleration bias of approximately
$10^{-8}\,{\rm m/s^2}$, one order of magnitude larger than the Pioneer
anomaly. Since the orientation of the solar arrays during the hibernation
phase is not actively maintained, a large uncertainty in the solar
radiation force on the spacecraft, $\sim 10^{-9}\,{\rm m/s^2}$, will
result. Hence both, the large unknown acceleration systematics and
the lack of regular tracking passes, will prohibit a test of the
anomaly with Rosetta.

Close to the class of exploration missions discussed in this work, is
NASA's New Horizons mission.\cite{stern} The destination of this
mission is Pluto and the launch is scheduled for 2006. Also for this
mission no test of the Pioneer anomaly is foreseen. On the contrary,
the mission baseline foresees that the spacecraft will be in a
spin-stabilised mode with little on-board activity and infrequent
tracking passes during most of the journey, similar to Rosetta. In
contrast to Rosetta this mode is not required by power constraints and
was mainly chosen to increase component lifetime and reduce operation
costs. Hence an enhanced tracking of the mission for a test of the
Pioneer anomaly would be possible in principle.  However doubts remain
that a sufficient knowledge of onboard acceleration biases can be
achieved to render such a test reliable. The system design of the
mission is far from ideal for a test of the Pioneer anomaly. The RTG
of New Horizons is directly attached to the spacecraft bus. This
design will lead to a considerable back-scattering of RTG heat from the
back of the antenna causing a large acceleration bias -- most likely
one order of magnitude bigger than the Pioneer anomaly -- along the
spin axis of the spacecraft. The determination of this acceleration
bias to sufficient accuracy in order to disentangle it from a putative
anomaly would most likely require a purpose made high-accuracy thermal
radiation model. The difficulties in the determination of the bias are
aggravated by a possible degradation of the surface properties of the
RTG and the back of the antenna during the flight. (see below
Sec. \ref{des}.\ref{desc} for a general discussion of these
issues). Hence, even with an enhanced tracking coverage, the system
design of the New Horizons spacecraft will be a considerable obstacle
for any attempt to verify the Pioneer anomaly with this mission.

The inability of various missions to achieve a long-term navigational
accuracy comparable to that of Pioneer~10 and 11 demonstrates that
both the system design and the trajectory design will need careful
consideration to accomplish a test of the Pioneer anomaly. From the
failure of Galileo and Ulysses and the deficits of New Horizons it is
clear that simply requiring a spin stabilised spacecraft on a mission
to the outer Solar System will not be sufficient.
Detailed considerations are necessary
to reduce the acceleration systematics on the test spacecraft to a
sufficient level.  In the
next section we will turn to the system design challenges posed by a
Pioneer anomaly test and we will present design solutions to reduce
the acceleration uncertainty that are feasible in nondedicated
scenarios.

\section{Nondedicated mission concepts \label{nonded}}

\subsection{The capabilities of exploration missions \label{capa}}

Dedicated missions to verify and characterise the Pioneer anomaly are
presently being intensively considered and at least two promising
concepts have been identified. The more conventional one is that of a
highly symmetric spacecraft with strong suppression of systematic
accelerations.\cite{gravitycontrol,Anderson:2002yc,Nieto:2003rq} The
acceleration sensitivity is expected to reach $10^{-11}\,{\rm m/s^2}$.
The performance of a Pioneer anomaly test is even further improved in
the second concept.\cite{Dittus:2005re,penanen} Here a
spacecraft with small acceleration systematics is envisaged to release
a small sub-satellite of ``reflective sphere'' type.  The sub-satellite
is completely passive is practically free of any systematic
accelerations. It is tracked from the mother-craft by laser ranging or
radar. The inter-spacecraft tracking signal is combined with radio
tracking of the mothercraft from the Earth to monitor any deviation of
the sub-satellite from geodesic motion. The expected acceleration
sensitivity of this setup is $10^{-12}\,{\rm m/s^2}$.

A nondedicated mission is not expected to reach the full performance
of the dedicated concepts. It has however the major advantage of
coming at considerably reduced costs provided a suitable mission can
be identified to host the experiment without interfering with the
primary mission goals.  Exploration missions to the outer Solar System
offer such an opportunity to test the Pioneer anomaly. Missions to
Uranus, Neptune or Pluto would most naturally feature, at a certain
point, a Jupiter gravity assist followed by a hyperbolic coast
arc. This coast phase lends itself to precision tracking of the
spacecraft trajectory which can be analysed to detect anomalous
accelerations. The major design drivers for such a mission would,
however, be the planetary exploration goals. Hence a design such as
the symmetric spacecraft described by Anderson et
al.\cite{Anderson:2002yc} would be excluded because of payload
requirements, e.\,g.\ field of view, and the need to accommodate a
propulsion module capable of achieving a capture into the orbit of the
target planet. The use of a special experimental payload able to test
the Pioneer anomaly test would most probably also be excluded because
of mass constraints. However, even under these conditions, an
investigation of the Pioneer anomaly is still attainable. Although
additional requirements on the spacecraft design are imposed, these
requirements can be fulfilled with no additional mass, little-to-no
impact on the other observational programme of the satellite, and no
additional risks.

We will first consider a class of low-mass, low-thrust missions
inspired by the study of a Pluto orbiter probe,
POP,\cite{POP1,Rathke:2004gv} and demonstrate the feasibility of a
Pioneer anomaly test on such a mission.
We then consider large spacecraft with electric propulsion powered
by nuclear reactors such as those sometimes envisaged to explore the
moons of the giant planets Jupiter and Saturn. One such spacecraft
was until recently considered by NASA under the name of Jupiter Icy
Moons Orbiter, JIMO. While the large amount of heat radiated from
the nuclear reactor on the craft would prohibit a test of the
Pioneer anomaly on the main spacecraft, this class of missions could
accommodate a small daughter spacecraft of less than 200\,kg mass
(Compared with the 1500\,kg of payload envisaged for JIMO). This
spacecraft could then be jettisoned during the approach of the
mothercraft to the target planet, and could use the planet for a
powered gravity-assist to achieve a ballistic hyperbolic trajectory.
The Pioneer anomaly test would then be performed by the daughter
spacecraft.

\subsection{The POP spacecraft \label{pop}}

Pluto Orbiter Probe (POP) is an advanced spacecraft designed within
the Advanced Concepts Team of ESA,\cite{POP1,Rathke:2004gv,POP2,Izzo}
that is capable of putting a 20\,kg payload into a low-altitude Pluto
orbit. The preliminary design has a dry mass of 516\,kg and a wet mass
of 837\,kg. The spacecraft is powered by four RTGs. The original
mission profile envisages a launch in 2016 and arrival at Pluto after
18 years of travel time, including a Jupiter gravity assist in 2018. A
suitable launch vehicle would be an Ariane~5 Initiative 2010. The
preliminary design of POP consists of a cylindrical bus, of 1.85\,m
length and 1.2\,m diameter. The Ka-band antenna of 2.5\,m diameter is
mounted on one end of the main structure. The four
general-purpose-heat-source (GPHS) RTGs are placed at the other end of
the main structure, inclined $45$\,deg to the symmetry axis of the
craft. The 4 QinetiQ T5 main engines are as well placed at this end of
the main structure. Next to the main engines in the main structure is
the propellant tank accommodating 270\,kg of Xenon propellant. POP is
a good example of what an advanced spacecraft to the outer Solar
System may look like and we therefore take it as a paradigm for this
kind of mission. Table \ref{scdata} gives the key figures of the
Planetary Orbiter Probe, that are relevant for our reasoning.

\subsection{The piggyback micro spacecraft  \label{pig}}

In the framework of NASA's Prometheus Program, JIMO was proposed by
NASA as the first mission to demonstrate the capabilities of
electric propulsion powered by a nuclear reactor. The mission,
recently cancelled in view of the new recent NASA priorities, is
still a plausible architecture for other future exploration mission.
Due to its high payload capabilities, a JIMO type of mission could
carry a micro spacecraft to test the Pioneer anomaly. The spacecraft
would be jettisoned at some point on the trajectory, and put into
hyperbolic heliocentric trajectory via a planetary gravity assist.
This would allow the spacecraft to perform a Pioneer anomaly test
after its swingby.

A possible baseline design for the piggyback spacecraft, resulting
from the design-driver of reducing on-board generated systematics, is
that of a spin-stabilised craft.  A preliminary mass estimate and
power budget can be based on the results of ESA's study of an
Interstellar Heliopause Probe,\cite{Lyngvi} which has a similar
baseline.  The result yields a mass of 150\,kg. The spacecraft would
use ion thrusters (e.\,g.\ hollow-cathode thrusters) for
attitude-control, and carry only a minimal scientific payload. Since
only a small data rate would be required, a 1.5\,m high-gain antenna
would be sufficient even in the outer Solar System. The required 80\,W
of power to operate the payload, the communication subsystem and the
AOCS (Attitude and On board Control System) would be provided by two
RTGs weighing 12.5\,kg each. Heat pipes from the RTGs to the main
structure of the spacecraft would be used for thermal control.

In addition,
a chemical propulsion module would be necessary to provide a
moderate $\Delta V$ before and during the swingby. This propulsion
stage would be jettisoned after the swingby, to eliminate the danger
that residual fuel might leak from the module and spoil the
Pioneer anomaly test. The dry mass of the module is estimated to be
16\,kg. A detailed design is beyond the scope of this article. We
apply a 20\% mass margin and a 20\% margin on the required power.
Accelerations due to on-board generated systematic errors are
inversely proportional to the mass of the spacecraft. Hence for the
calculation of the error budget, the conservative estimate will
arise from assuming the lower mass for the spacecraft but the higher
power consumption. The relevant parameters considered for the
piggyback micro spacecraft are summarised in Table\ \ref{scdata}.
\begin{table}[ht]
\begin{center}
\begin{tabular}{l@{\hspace{0.5cm}}r@{\hspace{0.5cm}}r}
\hline\hline
&POP&micro spacecraft\\
\hline
wet mass during coast $/$ kg & 750& 150\\
electric power $/$ W & 1000 & 100\\
RTG heat $/$ W & 10000& 1000\\
maximal radio-transmission power $/$ W& 50&10\\
antenna diameter $/$m&2.5 &1.5\\
\hline\hline
\end{tabular}
\caption{Overview of relevant spacecraft data for the two mission
  paradigms
\label{scdata}}
\end{center}
\end{table}

\section{Spacecraft design \label{des}}

From our review of missions to the outer Solar System we saw that a
 major obstacle for a test of the Pioneer anomaly is a lack of
 knowledge about the onboard generated forces, which are typically one
 order of magnitude larger than the Pioneer anomaly (cf.\ Longuski et
 al.\cite{Longuski}).  The aim of this section is to demonstrate that
 it is possible to reduce the overall on-board generated systematics
 to less than $10^{10}\,{\rm m/s^2}$, i.\,e.\ less than 10\% of the
 Pioneer anomaly by adopting, at the early design phase, some
 spacecraft design expedients that do not spoil the planetary-science
 mission objectives. We review the major possible sources of
 systematics and discuss how to reduce them to an acceptable level by
 a careful system design. Emphasis will be put on proof of
 concept by analytical considerations, that facilitate physical insight into
 the proposed methods.

\subsection{Thrust history uncertainties \label{desa}}

A precise knowledge of the thrust history of the spacecraft is
necessary if we want to be able to see small forces acting on the
spacecraft.\cite{Longuski}. However the thrust level of chemical or
cold-gas control thrusters varies considerably from firing to
firing. On top of this, the firing of a thruster is usually followed
by a considerable ``non-propulsive'' outflow of propellant, which
generates accelerations easily exceeding the magnitude of the Pioneer
anomalous acceleration (see Anderson et al.\cite{Anderson:2001sg}). A
more precise thrust history becomes available if ion engines are used
for the control of the spacecraft.  In addition, electric-propulsion
systems generate considerably smaller forces by nonpropulsive fuel
outflow (see Sec. \ref{des}.\ref{desb} below).

A more efficient solution is to reduce the number of attitude control
manoeuvres. This is achieved by spin stabilisation of the
satellite. For the piggyback micro-spacecraft paradigm this poses no
problems, and it is convenient to choose a relatively high rotational
velocity in order to guarantee the highest possible stability against
disturbances. For the POP paradigm, spin stabilisation seems to be in
contradiction to the requirements of planetary science, as the
instruments for the latter require high pointing accuracy, pointing
stability and slew rate capabilities that are not provided by a
spin-stabilised spacecraft. In reality the requirements of a
Pioneer-anomaly test and planetary science are not in contradiction as
the different objectives have to be fulfilled in different parts of
the mission. Hence the spacecraft can be in spin stabilised mode
during the coast phase, which will be used for the search for new
forces, and change to 3-axis stabilised mode when approaching its
final destination. Also, for any gravity assist 3-axis stabilised
control is desirable as it allows for a more precise control of the
nominal swingby trajectory. The spin-up before and spin-down after the
coast, in which the anomaly is tested, will be performed in deep
space, where few external disturbances act on the spacecraft. Hence
the spin-up and spin-down can be conducted over a long time span and
will only consume a negligible amount of propellant (see Izzo et
al.\cite{Izzo}).  Furthermore no additional attitude acquisition
hardware will be necessary. Thanks to the low disturbance level in
deep space, the rotational velocity of the satellite can be very low,
$\sim 0.01\,{\rm rpm}$, and the star trackers for the 3-axis
stabilised mode would still be sufficient for attitude
acquisition. Indeed, the coast in spin-stabilised mode might even save
mass, because it reduces the operating time of the momentum/reaction
wheels or gyros, and hence reduces the required level of redundancy.

\subsection{Fuel leaks and out-gassing \label{desb}}

A fuel leak from the attitude control system presents one of the
best candidates for a conventional explanation of the Pioneer
anomaly. Unfortunately, even in a new mission, it would be difficult
to entirely eliminate the possibility of fuel leaks caused by a
malfunctioning valve. The force $F$ generated by a mass flow rate
$\dot{m}$ is given by (see Longuski et al.\cite{Longuski}):
\begin{equation}
F = \dot{m} \sqrt{ 2 R T_s \left( 1+\kappa \right) / \kappa } \, .
\nn
\end{equation}
For chemical propellant systems the stagnation pressure corresponds to
the temperature in the tank $T_s = T_\text{tank}$\,. Requiring that
the maximal additional acceleration generated by propellant leakage
should not exceed $10^{-11}\,{\rm m/s^2}$, that is, remains two orders
of magnitude below the anomaly, then the maximally allowed
forces are $F \lesssim 7.5 \times 10^{-9}\,$N for the POP scenario and
$F \lesssim 1 \times 10^{-9}\,$N for the micro spacecraft.  The
corresponding mass-flow rates allowed would therefore be less than
5\,g/year, assuming realistic tank temperatures higher than 100K. This
requirement is far too demanding for a typical chemical attitude
control system (see\ Longuski and K\"onig\cite{Longuski}).

The problem of fuel leakage becomes more manageable for electric
propulsion systems, which do not rely on high tank pressures to
generate additional thrust. The propellant gas passes from the
high-pressure tank at $\sim150$\,bar and $\sim300$\,K, through a
central pressure regulator, before it is distributed to the engines
at low pressure, $\sim 2$\,bar. A redundant layout of the pressure
regulator would thus considerably reduce  the risk of leakage
by a valve failure. The internal leakage rate of a central
pressure regulator in an electric engine piping is typically
(assuming Xenon as a fuel without loss of generality) $\sim
10^{-8}\, {\rm lbar/s}$, 
and the external leakage is approximately $10^{-12}\, {\rm
lbar/s}$. From these numbers it is clear, that while external leakage
is sufficiently under control for the purpose of a Pioneer anomaly
test, it is necessary to further reduce internal leakage. This can
be achieved by placing a small reservoir with a low-pressure valve
after the central regulator. For the low pressure valve an even
smaller internal leakage is attainable, while the reservoir
accommodates the gas leaking through the regulator until the next
thruster firing, so that pressure build-up infront of the low-pressure
valve stays within its operational range. Hence, the use of electric
propulsion as an attitude control system alleviates the problem of
fuel leaks, and one of the major candidates of systematics on the
Pioneer probes can be eliminated, allowing us to assume $\Delta
a_\text{leak} = 10^{-11} \, {\rm m/s^2}$ for both mission concepts
under consideration.

Outgassing from the main structure of the spacecraft will, in general,
not play a big role in the error budget. This is mainly due to the
fact that the probe will already have traveled for a considerable time
before the test of the Pioneer anomaly will be performed.  Nearly all
outgassing will have taken place when the probe was closer to the
Sun. A more important source of outgassing are the RTGs of the
spacecraft. In general the $\alpha$-decay reaction in RTGs produces
helium, which will evaporate from the spacecraft. The decay of 1\,kg
of $^{238}$Pu produces approximately $4.2 \times 10^{-12}$\,kg/s of
helium. Assuming an efficiency of $40\,{\rm W/kg}$ for the generation
of electrical power (e.\,g.\ 38.3\,W$/$kg for the GPHS RTG used on
Cassini), we obtain a helium flow rate per generated watt of electric
power of $\dot{M}_\alpha / P = 1.1 \times 10^{-13}\,{\rm
kg/Ws}$. Furthermore it is reasonable to assume that the helium has
reached thermal equilibrium before it flows out of the
RTGs.\footnote{Actually, the helium plays an important role for
thermal conduction in the RTG. We are grateful to M.\ M.\ Nieto for
this information.} Then its average velocity is given by $v_\alpha =
\sqrt{3 k T/m_\alpha}$, where, $m_\alpha$ is the mass of a helium atom
and the temperature of the RTG will typically be about $T=
500\,$K. Hence the out-stream velocity of the helium will be $v_\alpha
= 1.7\times 10^3\,{\rm m/s}$. Assuming that all helium flows out
unidirectionally, and taking into account the power and mass values
given in Table \ref{scdata}, we may work out the magnitude of the
acceleration for the two spacecraft designs. In particular for
missions that have a nuclear electric propulsion system, the expulsion
of helium can make an important contribution, and its recoil effect on
the spacecraft needs to be taken into consideration. This is done most
easily by placing the pressure relief valves of the RTGs in a
direction that no net force results along the spacecraft's spin-axis.
The measure is particularly convenient because it does not require any
modification of the RTG design. We assume that the uncertainty in the
acceleration due to helium outgassing can be constrained to $2\%$ of
its worst case value, which corresponds to a placement of the valve
perpendicular to the spin axis with a precision of $1\,$deg. For the
planetary exploration missions this results into an uncertainty of
$\sim 4.2 \times 10^{-12}\,{\rm m/s^2}$ and for the piggyback concept
we find an uncertainty of $\sim 2.1 \times 10^{-12}\,{\rm m/s^2}$.

\subsection{Heat \label{desc}}

Heat is produced and radiated from the spacecraft at various points.
The dispersion of heat, necessary to maintain the thermal equilibrium
in the spacecraft, produces a net force on the spacecraft, of
$F = P_a/c =3\times10^{-9}\,$N per Watt of non isotropically radiated heat.

The heat generated in the main structure of the spacecraft will, in
general, be of the order of a few 100\,W. Assuming the above
advocated spin stabilisation of the craft, the thermal radiation
perpendicular to the spin axis of the satellite will average out
over one rotation. Hence the radial component of thermal radiation
does not contribute to the error budget for the measurement of a
putative near-constant, i.\,e.\ very low-frequency, acceleration. By
placing the radiators so that the heat they dissipate does not
produce a net force along the spacecraft axis, the contribution of
the thermal radiation force can be reduced to a few watts. Note that
the avoidance of reflections is much superior to the precision
modelling of the thermal radiation characteristics of the spacecraft
because the effect of surface deterioration during the journey is
difficult to model. Thus, the avoidance of reflections by restricting
the opening angle of radiators is mandatory for a precision test of
the Pioneer anomaly. The radiation from other surfaces of the
spacecraft can be monitored to some extent by measurements of the
surface temperature. This option is discussed below for the case of
the RTGs. We will therefore assume as a spacecraft design
requirement that radiators are positioned in such a way as to reduce
the total force due to the radiated heat along the spacecraft spin
axis to a fraction of the Pioneer anomaly. We will set $\Delta
a_{\rm bus} = 1\times 10^{-11}\,{\rm m/s^2}$.

By far the bigger source of thermal radiation are the RTGs, necessary
to power the spacecraft systems. In particular if one chooses an
electrical propulsion system, the thermal heat to be dissipated from
the RTGs may easily reach 10\,kW for the exploration
paradigm.\cite{POP1} In principle an anomaly caused by RTG heat can be
distinguished from other sources because it will exponentially decay
with the 87.7 years of half-life of the plutonium, which would result
in a change of approximately 8\% in 10 years. In the case of the
Pioneer spacecraft however, the disturbances by attitude control
manoeuvres were so large that no reliable determination of a possible
slope of the anomaly could be performed.\cite{Markwardt} For a new
mission, in which gas leaks are well under control, a reliable
measurement should however be possible. Nevertheless it is desirable
to have an independent upper limit on the effect of RTG heat so that a
reliable estimate can be given of this effect for each interval
between attitude-control manoeuvres.

Hence it is preferable to reduce forces due to non-isotropic
heat-emission from the RTG to the level of a fraction of the
expected anomaly. To accomplish this, RTG heat must be dissipated
fore--aft symmetrically, and reflections from the spacecraft should
be avoided. This may be simply achieved by putting the RTGs on long
booms or reducing their view factor towards other components of the
spacecraft by a more intricate design. In combination with a
detailed model of the radiation characteristics, this reduces any
unmodelled directional heat radiation resulting from asymmetry to
affordable values.

More troublesome is the effect of possible material degradation on
the radiation characteristics of the RTGs. During a typical mission,
the antenna-facing side of the RTGs will be exposed to solar
radiation almost permanently, whereas the other side of the RTGs
lies in shadow for nearly all of the mission. Hence one can expect a
very asymmetric degradation of the emissivity and absorptance of the
RTGs. Whereas it would be difficult to predict which part of the RTG's
surface degrades faster -- most likely it would be the sun-facing
side -- one can reconstruct the overall degradation of the emissivity
$\epsilon$ of the RTG by monitoring its temperature $T$ at selected
points.

We demonstrate the reconstruction of acceleration originating from
degradation o f optical properties of the RTG for a simplified model
of a cylindrical RTG, with the cylinder axis perpendicular to the
spacecraft--Sun direction. As a further simplification we assume
perfect thermal conductivity of the RTG so that all of its surface is
at the same temperature. We first derive a relation between the
temperature and the emissivity change and then a relationship between
the resulting change in acceleration and the emissivity change. We
then show how under reasonable assumptions temperature and
acceleration can also be directly related.

The azimuth angle $\psi$ of the cylinder is measured from the
Sun-pointing direction. Using the Stefan-Boltzmann law, the relation
between the total radiated power, $P_\text{tot}$, the emissivity per
angle $\epsilon (\psi)=\epsilon_0 + \Delta \epsilon (\psi)$ and the
temperature of the RTG is given by
$$
P_\text{tot} = \frac{{\rm const.}\, T^4}{2\pi \epsilon_0}
\int_0^{2\pi} [\epsilon_0 + \Delta\epsilon (\psi)] d \psi \, .
$$
Since the thermal power produced by the RTG is well known from the
amount of plutonium in it, the temperature of the RTG is directly
related to change of emissivity $\Delta\epsilon$. Indicating with
$T_0$ the temperature of the RTG when $\Delta\epsilon(\psi)=0$, we
have:
$$
T = T_0 \left( \frac{2\pi \epsilon_0} {2\pi\epsilon_0 +
\int_0^{2\pi} \Delta \epsilon (\psi)d \psi } \right)^{1/4} \, .
$$
On the other hand the power per angle is related to the total radiated power
by
\begin{equation}
P(\psi) = P_\text{tot} \frac{\epsilon_0 + \Delta \epsilon
(\psi)}{\int [ \epsilon_0 + \Delta \epsilon (\psi) ] \, d \psi} \, .
\label{Pofphi}
\end{equation}
The effective asymmetric power radiated along the spin axis of the
craft is given by
\begin{equation}
P_a = \int_0^{2\pi} P(\psi) \cos (\psi ) \, d\psi \, .\
\label{asyP}
\end{equation}
Inserting Eq.~(\ref{Pofphi}) into Eq.~(\ref{asyP}) and expressing
the acceleration $a_\epsilon$ induced by the change in emissivity,
we obtain
\begin{equation}
a_\epsilon = \frac{P_\text{tot}}{M_\text{S/C} c} \frac 1{\int
[ \epsilon_0 + \Delta \epsilon (\psi) ]\,  d \psi}
\int_0^{2 \pi} \cos (\psi)
[ \epsilon_0 + \Delta \epsilon (\psi) ]
\, d \psi \,  .
\label{aeps}
\end{equation}
In general there will be no unique relation between $T$ and
$a_\epsilon$ because the quantities depend on different integrated
functions of the emissivity.  Nevertheless a relation can be
established under some reasonable model assumptions. To
illustrate this we consider an RTG which has an original emissivity
of $\epsilon_0 = 1$, and we model the emissivity change with the
simple relation:
\begin{equation}
\Delta \epsilon (\psi) = - \Delta \epsilon_\text{max} \cos (\psi)
\text{\,,~~for~} |\psi| \le \pi/2 \, ,
\nn
\end{equation}
where $\Delta \epsilon_\text{max} > 0$ is the absolute value of the
change of emissivity in the Sun-pointing direction. In this case
the deceleration of the spacecraft is given by
\begin{equation}
a_\epsilon  = - \frac{P_\text{tot}}{4 M_\text{S/C} c}
\Delta \epsilon_\text{max}\, .
\nn
\end{equation}
The temperature after the degradation of emissivity is then related to
the temperature at nominal emissivity $T_0$,
\begin{equation}
T = T_0 + \frac{T_0}{4 \pi}\, \Delta \epsilon_\text{max} + O
(\Delta\epsilon_\text{max}^2) \, . \nn
\end{equation}
We obtain the final relation
\begin{equation}
a_\epsilon=-\frac{\pi P_\text{tot}}{c M_{S/C}} \frac{T-T_0}{T_0} \, .
\label{eq:AccelTempRelation}
\end{equation}
Consequently we find for the acceleration uncertainty $\Delta
a_\epsilon$ a dependence on the temperature uncertainty $\Delta T$
\begin{equation}
\Delta a_\epsilon=\frac{\pi P_\text{tot}}{c M_{S/C}} \frac{\Delta
T}{T_0} \, .  \nn
\end{equation}
 For an RTG the nominal temperature is $T_0 \sim 500\,$K.
Hence, assuming that we monitor the RTG temperature at a precision
of $0.1\,$K and assuming the above degradation model, we would have
an uncertainty in the anomalous acceleration of $\Delta a_\epsilon =
2.8\times 10^{-11}\,{\rm m/s^2}$ for the exploration scenario and
$\Delta a_\epsilon = 1.4\times 10^{-11}\,{\rm m/s^2}$ for the
piggyback scenario.

A realistic model of the RTG is considerably more complicated. It
has to include the absorptance, and to account for a non-uniform
temperature of the RTG and the Yarkovsky effect (see\
Cruikshank\cite{Yarkovsky} or Bottke et al.\cite{Bottke}). These are,
however, mainly numerical complications, and it is always possible
to develop a refined version of Eq.~(\ref{eq:AccelTempRelation}) so
that the uncertainty of the RTG temperature measurements may be
related to the uncertainty of the derived acceleration. In
particular there is no danger of mistaking a degradation or failure
of thermocouples of the RTG for a change in emissivity because these
effects are distinguishable by the accompanying decrease of electric
power. Hence we assume that the acceleration levels
found in the simple model are also achievable in a realistic situation.

\subsection{Radio-beam radiation force \label{desd}}

The increasing amount of data gathered by modern planetary observation
instruments demands high data transmission capabilities.  For
exploration missions to the outer Solar System like the ones discussed
here this inevitably leads to high transmission powers for the
telecommunication system, $P_\text{Radio} \sim 50$\,W. Analogous to
the case of thermal radiation, discussed in the previous section, the
acceleration on the spacecraft is given by
\begin{equation}
\vett a_\text{Radio} \approx \frac{P_\text{Radio}}{cM_{S/C}} \vett e_A
\nonumber
\end{equation}
in the approximation of a narrow radio beam. 
Hence it may easily reach the order of magnitude
of the Pioneer anomaly. However this bias can be constrained
in a straightforward way. During the coast phase, in which the
Pioneer anomaly is to be tested, the data volume generated on-board
will be much smaller than at the final destination of the probe.
Hence the transmission power can be reduced to a few Watts during
the test, bringing the uncertainty in the transmission power for
both mission paradigms down to less than 1\,W. This would correspond
to an acceleration systematics below $\Delta a_{\text{Radio}} = 5
\times 10^{-12}\,{\rm m/s^2}$ for the planetary exploration mission
and $\Delta a_{\text{Radio}} = 2.2 \times 10^{-11}\,{\rm m/s^2}$ for
the micro spacecraft. These numbers might be even further
reduced by changing the transmission power to different values
during the measurement period and measuring the subsequent change of
the spacecraft acceleration. In this way one could actually
calibrate for the effect of the radiation beam.

\subsection{Solar radiation pressure \label{dese}}

The last major contribution to discuss in this context is the solar
radiation pressure. For the present level of analysis it is
sufficient to discuss the effect of the solar radiation force by
considering the force on a flat disk of the size of the spacecraft
antennas and covered with white silicate paint. To further simplify
our consideration we restrict ourselves to specular reflection and
neglect diffuse reflection and the Yarkovsky effect.  Then we can
express the acceleration induced by solar radiation
pressure as\cite{vanderha}
\begin{equation}
\vett a_\odot = \frac{P_\odot}{c\,r^2} \frac{A_{S/C}}{M_{S/C}}
(1 + \eta) \cos^2 \theta \, \vett e_A
\, ,
\label{asolar}
\end{equation}
where we have used the fact that the tangential force arising from
the partial specular reflection has no effect on the centre-of-mass
motion of the spacecraft due to the spin stabilisation. 
Since the antenna is oriented towards Earth, the vector $\vec{e}_A$ is
Earth-pointing, and the two vectors $\vett e_A$ and $\vett e_\odot$
only enclose a small angle $\theta$ for large heliocentric distances,
i.\,e.\ in all mission options for most of the measurement phase (see
below Sec.~\ref{traj}). The uncertainty of the acceleration due to solar radiation
$\Delta \vec{a}_\odot$ is dominated by a possible change in the
reflectivity properties of the spacecraft. We assume $\Delta
\eta/\eta_0 = 5\%$. The uncertainties of all other quantities are
about one order of magnitude smaller and can be neglected for our
purposes. Hence we find from Eq.\ (\ref{asolar}), the acceleration
uncertainty due to solar radiation pressure
\begin{equation}
\Delta \vett a_\odot =
\frac{P_\odot}{c\,r^2} \frac{A_{S/C}}{M_{S/C}} \cos^2 \theta \vett e_A
\, \Delta \eta
\label{solarpres} \, .
\end{equation}
The maximal value is taken for $\cos \theta = 1$. For the planetary
exploration scenario we find $\Delta a_\odot=149\,(r_\oplus/r)^2 \times
10^{-11}\,{\rm m/s^2}$ and for the piggyback concept $\Delta
a_\odot=268\,(r_\oplus/r)^2 \times 10^{-11}\,{\rm m/s^2}$. We see from
these numbers that the uncertainty on the solar radiation force model
would exceed one third of the putative anomaly at heliocentric
distances of less than 3\,AU for the micro spacecraft and 2\,AU for the
exploration mission. Below these heliocentric distances a reliable
detection of the anomaly would become impossible.

\subsection{Summary of the onboard error sources \label{err}}

In the previous section we discussed the major sources of systematic
effects on the spacecraft acceleration for the two nondedicated
concepts under consideration and we have determined the uncertainties
to which they can be restricted by suitable design measures.
 The numerical results are
summarised in Table \ref{budgetov}. For a spin-stabilised craft, all
acceleration uncertainties arise along the rotational axis of the
spacecraft, with the exception of the solar radiation pressure.

\begin{table}[t]
\begin{center}
\begin{tabular}{l@{\hspace{0.5cm}}r@{\hspace{0.5cm}}r}
\hline\hline
Source of& POP paradigm & micro spacecraft\\
acceleration
&$\Delta a / (10^{-11} \rm{m/s^2})$ &$\Delta a / (10^{-11} \rm{m/s^2})$\\
\hline
Fuel leaks& 0.4&0.2\\
Heat from bus& 1.0&1.0\\
Heat from RTG & 2.8&1.4\\
RTG helium outgassing& 2.7& 2.0\\
Radio beam& 0.5&2.2\\
Solar radiation pressure& $149\,(r_\oplus/r)^2 \cos^2 \theta$&
$268\,(r_\oplus/r)^2 \cos^2 \theta$\\
\hline
Total  & $7.4 + 149\,(r_\oplus/r)^2 \cos^2 \theta$ 
& $6.8 +268\,(r_\oplus/r)^2 \cos^2 \theta$  \\
\hline\hline
\end{tabular}
\caption{Acceleration uncertainties for the two mission
  paradigms.
\label{budgetov}}
\end{center}
\end{table}

The sources of acceleration, which were identified are uncorrelated --
at least to the level of the modelling performed -- and the overall
acceleration due to systematics is therefore bounded by the value
$\Delta a = \sum_i\Delta a_i$. This returns $\Delta a = [ 7.4 +
149\,(r_\oplus/r)^2 \cos^2 \theta ]\,\times 10^{-11}{\rm m/s^2}$for the
exploration mission and $\Delta a = [6.8 + 268\,(r_\oplus/r)^2 \cos^2
\theta ]\, \times 10^{-11}{\rm m/s^2}$ for the piggyback micro
spacecraft. This would only, when sufficiently far from the Sun, allow
determination of the anomaly to a precision of 10\,\%, which is
approximately one order of magnitude worse than the error-budget
presented by Nieto and Turyshev\cite{Nieto:2003rq} for a highly
symmetric dedicated spacecraft.

The accuracy to which an anomalous acceleration can be determined
will also strongly depend on its direction. Since all error sources
will cause an acceleration purely along the spin axis of the
spacecraft, they will be competing with an Earth-pointing anomaly,
which would most likely be an effect on the radio signal. When
studying the capabilities of the mission to discriminate the
direction of the anomaly, the systematic errors do not influence the
result because their direction does not change and their magnitude
has a gradient, which cannot be confused with a direction-dependent
modulation.

\subsection{Summary of spacecraft design \label{dessum}}

From the goal to minimise the uncertainties in conventional
accelerations, we have arrived at several design requirements for our
spacecraft: Spin stabilisation of the spacecraft seems mandatory in
order to reduce the number of attitude control manoeuvres of the
spacecraft. Furthermore it ensures that all onboard-generated
accelerations are pointing along the spin axis of the craft. This
effectively eliminates the effect of systematics on the determination
of the direction of a putative anomaly (see below
Sec. \ref{trac}.\ref{perf}). For the exploration scenario spin
stabilisation is most practically only chosen during the coast phases
of the mission. An electric propulsion system turns out to be the most
promising option to reduce the amount of acceleration systematics from
propellant leakage, although an electric propulsion system has the
disadvantage, that due to its high power consumption it considerably
increases the amount of heat generated on board the spacecraft. The
major source of asymmetric thermal radiation from the craft are the
RTGs. The heat systematics can be constrained to a sufficient degree
by monitoring the temperature of the RTGs. Furthermore the view factor
of the RTGs from the spacecraft bus and the antenna should be made as
small as possible in order to reduce radiation back-scattering and
simplify the modelling. In order to constrain the systematics induced
by the radio transmission beam the transmission power during the
measurement phase can be reduced to a few Watts.

While the requirements imposed on the spacecraft make it necessary
that the spacecraft is already designed with the goal of testing the
Pioneer anomaly under consideration, the modifications suggested come
at no increase in launch mass and at no increase in risk. In
particular, the goal of testing the Pioneer anomaly is compatible with
the constraints of a planetary exploration mission.  The sources of
acceleration uncertainty and counter-measures that have been discussed
in this Section are summarized in Table~\ref{designov}.

\begin{table}[t]
\begin{center}
\begin{tabular}{l@{\hspace{0.5cm}}p{9cm}}
\hline\hline
Source of acceleration uncertainty & Suggested counter measure \\
\hline
Thrust history uncertainty & Spin stabilization\\
Fuel leaks& Electric propulsion\\
Heat from spacecraft bus& Placement of radiators, spin stabilisation\\
Heat from RTGs & Reconstruction from monitoring of RTG temperature\\
RTG helium outgassing& orientation of pressure relief valves on RTG's\\
Radio beam force& Low transmission power during test\\
Solar radiation pressure& Sufficient heliocentric distance \\
\hline\hline
\end{tabular}
\caption{Sources of acceleration uncertainties and possible design solutions.
\label{designov}}
\end{center}
\end{table}

\section{Measurement strategies \label{trac}}

\subsection{Instrumentation options \label{inst}}

A mission to test the Pioneer anomaly has to provide three types of
information. It must monitor the behaviour of the tracking signal
for an anomalous blueshift; it must be able to detect an anomalous
gravitational force acting on the spacecraft; and it must also be
capable of detecting an anomalous non-gravitational force on the
spacecraft. From these three tasks it is obvious that radio tracking is the 
experimental method of choice because it is sensitive to all three of the possible sources of the Pioneer anomaly. Radio tracking will be analysed extensively in the following sections \ref{trac}.\ref{meth} to \ref{trac}.\ref{perf}.

However the orbit reconstruction from radio tracking data does not
discriminate between a non-gravitational (systematics) and a
gravitational (new physics) origin of the anomaly. Such conclusions
can only be drawn from a statistical test of a specific candidate
model against the observed deviation from the nominal orbit. Hence a
model-independent discrimination between a gravitational and
non-gravitational anomaly would be highly desirable. Such a
distinction could in principle be accomplished with an accelerometer
on board the spacecraft, because deviations of the spacecraft from a
geodesic motion will be induced by non-gravitational forces
only. Unfortunately the use of accelerometers reaching the sensitivity
level of the Pioneer anomaly is excluded by weight constraints:
high-precision accelerometer assemblies weigh typically in the order
of 100\,kg (cf.\ e.\,g.\ ESA's GOCE mission\cite{GOCE}). Concludingly, the
discrimination between a gravitational and non-gravitational anomaly
will have to rely on the interpretation of the tracking data.

In order to improve our understanding of disturbing forces generated
by the space environment in the outer Solar System, and to make sure
that they cannot contribute significantly to the Pioneer anomaly, it
is desirable to include a diagnostics package in the payload,
consisting of a neutral and charged atom detector and a dust
analyser. The mass of such a package is approximately
1.5\,kg.\cite{Lyngvi} Most likely it would be part of the payload due
to space science interests anyway as in the case of the New Horizons
mission.\cite{stern}

\subsection{Tracking methods \label{meth}}

We briefly review the available tracking methods to
explain how their combination allows an unambiguous discrimination
between the various possible causes of the anomaly (see Thornton and
Border\cite{Thornton} for an introduction to tracking methods).

In sequential ranging, a series of square waves is phase modulated
onto the uplink carrier. The spacecraft transponds this code.  The
ground station compares the transmitted and the received part of the
signal and determines the round-trip time from the comparison. Since
the modulated signal is recorded and compared in order to obtain the
distance from the spacecraft to the ground station, the information
obtained relies on the group velocity of the signal. The group
velocity is influenced by the interplanetary plasma, which acts as
dispersive medium, but not by gravitational effects, which are
non-dispersive. For this technique we assume a range error
$s_\text{track} = 0.6\,m$ at $1\sigma$ confidence level in our
analysis.\cite{Thornton}

Doppler tracking uses a monochromatic sinusoidal signal. The signal
is sent to the spacecraft and is coherently transponded back to
Earth. The phases of both the outgoing signal and the incoming
signal are recorded. Since the frequency of the wave is the
derivative of the phase, the frequency change between the outgoing
and incoming wave can be determined, and the relative velocity of
the spacecraft and the tracking station can be inferred. The
position is then obtained by integrating the observed velocity
changes to find the distance between the spacecraft and the tracking
station. The Doppler data are sensitive to other phase shifting
effects such as the frequency shift by the interplanetary plasma and
to a gravitational frequency shift. For a long integration time the
Doppler error is usually dominated by plasma noise, which typically
leads to an error of approximately $v_\text{track}=0.03\, {\rm
mm/s}$ at $1\sigma$ confidence level.\cite{Thornton,Kinman}

The simultaneous use of both tracking techniques allows for a
correction of charged medium effects, because for a signal that
propagates through a charged medium the phase velocity is increased
whereas the group velocity is decreased by the same
amount.\cite{Curkendall2} The comparison of the Doppler and ranging
measurements in order to determine plasma effects has important
benefits for nondedicated test of the Pioneer anomaly, because it
allows a determination of the errors induced by the charged
interplanetary medium without requiring dual frequency capabilities,
and is thus a considerable mass saver. Since the information of the
sequential ranging relies on the group velocity of the signal, and
the information of the Doppler tracking relies on the phase velocity
of the carrier, the use of both ranging methods also allows
distinction between a real acceleration of the spacecraft and an
anomalous blueshift. Whereas a real acceleration would show up in
both data, the frequency shift would only affect the Doppler signal,
which is sensitive to changes in the phase velocity of the wave but
not to the sequential ranging signal that measures the group
velocity.

Both Doppler tracking and sequential ranging are primarily sensitive
to the projection of the spacecraft orbit onto the Earth--spacecraft
direction. In order to characterise a putative anomaly it is however
crucial to determine its direction. In view of this problem, it
could be beneficial to obtain independent information on the motion
of the spacecraft orthogonal to the line of sight. This information
is in principle provided by Delta differential one-way ranging
($\Delta$DOR). Differential one-way ranging determines the angular
position of a spacecraft in the sky by measuring the runtime
difference of a signal from the spacecraft to two tracking stations
on Earth. Assuming that the rays from the spacecraft to the two
stations are parallel to each other, the angle between the
spacecraft direction and the baseline connecting the two stations
can be determined from the runtime difference. In $\Delta$DOR the
accuracy of this method is further improved by differencing the
observation of the spacecraft from that of an astronomical radio
source at a nearby position in the sky. The typical accuracy
achievable with $\Delta$DOR is $\alpha_\text{track}=50\,{\rm nrad}$ at
$1\sigma$ confidence level.\cite{Thornton} An improvement in
accuracy of two orders of magnitude in angular resolution would be
achievable if the next-generation radio-astronomical interferometer,
the Square Kilometre Array, could be used for the
tracking.\cite{Jones} However this observatory is not likely to be
completed by the launch dates under consideration. Hence we do not
include this enhanced capability in our analysis.

\subsection{Tracking observables for the Pioneer anomaly test \label{obs}}

The suitability of an interplanetary trajectory for a test of the
Pioneer anomaly may be influenced by a dependence of the Pioneer
anomaly on the orbital parameters of the trajectory, as already
discussed in Sec.~\ref{PA}.\ref{PAe} above.  The second important
criterion for the choice of trajectory is if it enable a precise
measurement of the properties of the anomaly. For the purpose of a
general survey of trajectory options for a broad class of missions a
simulation of the tracking performance for each trajectory becomes
unfeasible due to the large computational effort involved. Hence we
resort to the opposite route: In this section we derive a linearised
tracking model for the anomaly that neglects the back-reaction of the
anomaly on the orbital parameters of the trajectory. This
model allows to express the performance of the tracking
techniques for a specific trajectory as a function of the
heliocentric distance of the spacecraft and the flight angle only.
 
The capabilities of the three tracking techniques are
evaluated by determining after which time a detectable
deviation from the trajectory has accumulated. The perturbation on
the position vector is well described, for our purposes, by the
simple equation:
\begin{equation}
\label{eq:simpledynamic} {\ddot{\vett s}}^*=\vett a^* \, .
\end{equation}
where $\vett s^* = \vett r - \boldsymbol \rho$ is the difference
between the position $\vett r$ of a spacecraft not affected by the
anomaly and the position $\boldsymbol \rho$ of a spacecraft affected
by the anomalous acceleration $\vett a^*$. In fact we may write (see
e.\,g.\ Bate et al.\cite{Bate} pp.~390--392) the full equation of motion in the
form:
\begin{equation}
\label{eom}
{\ddot{\vett
s}}^*+\frac{\mu_\odot}{r^3}\left[\left(\frac{r}{\rho}\right)^3-1\right]\vett
r+\mu_\odot\frac{\vett s^*}{\rho^3}=\vett a^*
\end{equation} 
Note that this holds also for non-Keplerian $\vett r$ whenever the
non-gravitational modelled forces may be considered state-independent
(as is the case for the systematic accelerations considered in
Sec.~\ref{des}.\ref{err}). At Jupiter distance, it takes roughly three
months for the second and third terms of Eq.~(\ref{eom}) to grow
within two orders of magnitude of $\vett a^*$. The smallness of the
back-reaction on the orbital parameters is also the reason why it is
not possible to decide from the Pioneer Doppler data if the observed
anomaly is caused by an effect on the radio signal or a real
acceleration. Thus Eq.~(\ref{eq:simpledynamic}) and its solutions,
\begin{equation}
\vett v^* = \int_{t_0}^{t_1} \vett a^*(t') dt' \, ,\text{~~and~~~~}
\vett s^*= \iint\limits_{t_0}^{\quad t_1} \vett a^*(t') dt'
\end{equation}
can be used to estimate 
the deviation from the nominal trajectory caused by the anomaly.

\begin{figure}[t]
\centering{\includegraphics[height=10cm]{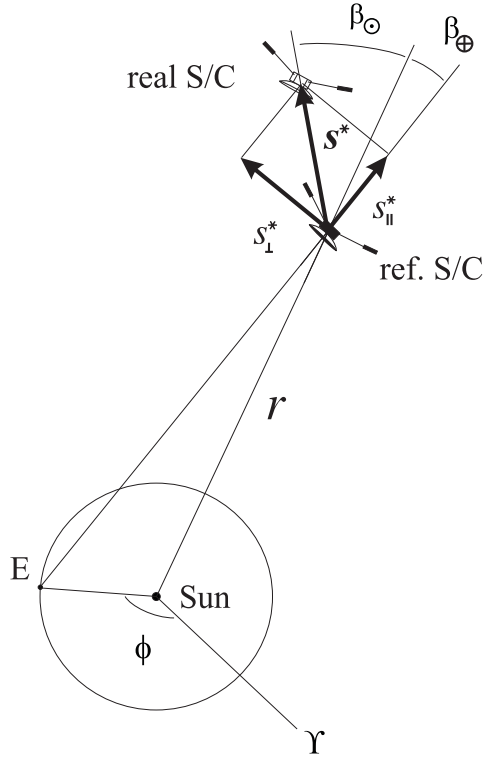}}
\caption{Tracking of the anomaly in the ecliptic plane.}
\label{geom}
\end{figure}

Without loss of generality we consider our spacecraft as lying in the
ecliptic plane. A depiction of the geometry for this two-dimensional
model is displayed in Fig.~\ref{geom}.  
Direct connection to the tracking observations in the
geocentric frame, in which the measurements are actually
conducted,\footnote{For our purpose we can neglect the purely technical
complications arising from the geocentric motion of the tracking
stations. See e.\,g.\ Montenbruck and Gill\cite{montenbruck} Chapter~5 for an
extensive discussion of this topic.} can be established by projecting
the anomalous velocity change and position change onto the
Earth--spacecraft vector. The change in the geocentric angular
position of the spacecraft in the sky, $\alpha_\oplus^*$, is obtained
from the component of $\vett s^*$ perpendicular to the
Earth--spacecraft direction through the relation $\alpha^*_\oplus
\simeq s^*_\bot / s$. We get
\begin{align}
v_\|^* &= |v^*(t_1)| \cos \beta(t_1) - |v^*(t_0)| \cos \beta(t_0) \, ,
\label{vpar} \\
s_\|^* &= |s^*(t_1)| \cos \beta(t_1) - |s^*(t_0)| \cos \beta(t_0) \, ,
\label{spar} \\
\alpha_\oplus^* &\simeq  \frac{1}{s(t_0)} \big[
|s^*(t_1)| \sin \beta(t_1) - |s^*(t_0)| \sin \beta(t_0)
\big]\, ,
\label{asc}
\end{align}
where $\beta$ is the angle between the anomaly direction and
the Earth--spacecraft vector. The equations (\ref{vpar})--(\ref{asc}) estimate
the effect of an anomalous acceleration on the tracking
observables. Note that the magnitude of the Doppler observable only
grows linearly with time, whereas the observables of ranging and
$\Delta$DOR grow quadratically in time.

It is convenient to express the angle $\beta$ as the sum of the angle
between $\vec{a}^*$ and the Sun--spacecraft vector $\beta_\odot$,
and the angle between the Earth--spacecraft vector and the
Sun--spacecraft vector $\beta_\oplus$ (see Fig.~\ref{geom}),
\begin{equation}
\beta = \beta_\odot + \beta_\oplus \, .\nn
\end{equation}
With this decomposition several relevant cases can easily be treated:
First is the case of a sun pointing acceleration from a central force,
$\beta_\odot \equiv 0$. Second is the case of an inertially fixed
acceleration $\beta_\odot = \text{const}$. This case also yields
insights into the case of a drag-force type deceleration along the
trajectory because the change of $\beta_\odot(t)$ stays typically
within the same magnitude as that of $\beta_\oplus (t)$. Third is the
case of a blueshift of light. It leads only to a change of the {\em
apparent} velocity along the line of sight of the spacecraft,
$v^*_{\|}$, but not of the position $\vett s^* \equiv 0$. From the
direction of the effect we have immediately $\beta \equiv 0$.

Both the rate of change $\dot{\beta}_\odot$ and the angle
$\beta_\oplus$ are small quantities for trajectories in the outer
Solar System. Hence analytical expressions for $\cos \beta$ and $\sin
\beta$ are conveniently obtained by expanding $\beta$ around the angel
at the begin of the tracking interval $\beta(t_0)$ in the quantities
$\dot \beta_\odot$ and the mean motion of the Earth.
After these steps the magnitude of the anomalous components of the
tracking observables in the above cases of special interest depends on
the heliocentric distance $r$ and the direction of the anomaly in the
heliocentric frame $\beta_\odot$, only. 

Expressions analogous to Eqs.~(\ref{vpar})--(\ref{asc}) hold for the
systematic acceleration uncertainties summarized in
Sec.~\ref{des}.\ref{err}.  Combining the expressions for the effect of
the anomaly and the systematic accelerations, one can determine the
sensitivity of a tracking technique to one of the three generic
classes of the Pioneer anomaly.

\subsection{Tracking performance \label{perf}}

We content ourselves with the evaluation of the tracking performance for an anomaly of constant magnitude as it is indicated by the Pioneer data.
The measurement performance is conveniently evaluated by splitting
the change in the generic tracking observable $f$ induced by the
anomaly into a constant component $f^*$ and a time-dependent
component, $\delta f^*$ which is dependent on the direction in which
the anomaly acts.

First we consider the detectability of the anomaly without attempting
to determine its direction. For this goal it is sufficient to consider
the zeroth order terms in the expansion around $\beta(t_0)$. In order
for the anomaly to be detectable the term $f^*$ has to exceed the
measurement error. The measurement error in turn is given by the sum
of the tracking error $f_\text{track}$ (we require a confidence level
of $3\sigma$) and the uncertainty in the systematic accelerations
$\Delta f$. The two errors have to be added instead of taking their
Pythagorean sum because the error induced by the uncertainty in the
systematic accelerations is not of statistic nature. Thus the
condition for detectability reads
\begin{equation}
f^* > f_\text{track} + \Delta f \,. 
\end{equation}

We first consider the case of Doppler tracking, $v^*_\| >
v_\text{track} + \Delta v \,$.
Solving for the tracking time we find
\begin{equation}
t_v > v_\text{track} / \left( a^* \cos \beta_\odot
- \Delta a  \right) \, .
\nn
\end{equation}
Proceeding analogously for sequential ranging and $\Delta$DOR
one finds that sequential ranging and Doppler tracking easily detect an
anomaly of $a^* \cos \beta_\odot = 10^{-9}\,{\rm m/s^2}$
within two days of tracking at heliocentric distances beyond Jupiter's orbit. 
The time to detection decreases slightly for larger heliocentric distance due to the decrease of the solar radiation pressure.
$\Delta$DOR cannot compete in
performance with the other tracking methods. Even if we consider
$a^* = 10^{-9}\,{\rm m/s^2}$ and $\beta_\odot$ as large as 30\,deg,
the detection of the anomaly takes 140 days even at 5\,AU and rises
to 370 days at 35\,AU.

Significantly more challenging than the detection of the anomaly is
the determination of its direction. Here we consider the detection
of the three most plausible candidate directions: Sun pointing,
along the velocity vector, and Earth pointing. The case of an
acceleration along the velocity vector is, to a good approximation,
covered by the case of an acceleration having a fixed angle with the
Sun--spacecraft vector, because the change of the flight angle of
the spacecraft along the trajectory will be very slow.

A Sun-pointing effect would be revealed by the variation of the
tracking observables due to the Earth's rotation around the Sun.  In
order to detect unambiguously the annual modulation $\delta f$ in a
tracking observable $f$, its modulation has to exceed the sum of the
tracking error $f_\text{track}$ and of the uncertainty in the annual
modulation of systematic accelerations $\delta (\Delta f)$. 
\begin{equation}
\left| \delta f^*_{\|} \right| > f_\text{track} + \left| \delta (\Delta f_\odot) \right|
\, .
\label{dtrack}
\end{equation}
The systematics term stems entirely from the uncertainty in the solar
radiation force, $\delta (\Delta f) = \delta (\Delta f_{\odot,\|})$
because all other accelerations are Earth-pointing and do not show a
modulation (cf.\ Sec.~\ref{des}.\ref{err}).  

We set an upper limit of 6 months on the detection time because this is
the expected approximate time span between two attitude control
manoeuvres. Longer time spans cannot be evaluated in search for the
modulation because the attitude manoeuvres are expected to
considerably degrade our knowledge of the orbital motion of the
spacecraft. Putting this limit on the observation time and using the
error budget determined in Sec.~\ref{des}.\ref{err}, we find that the
annual modulation is detectable by Doppler tracking up to 6.2\,AU
heliocentric distance for both the exploration spacecraft and for the
micro spacecraft.
Applying the same reasoning for sequential ranging, one finds that
for both paradigms the annual modulation remains detectable in
sequential ranging beyond 50\,AU.
For $\Delta$DOR the modulation term is suppressed compared with the
constant term by a factor of $r_\oplus/r$. Considering the poor
performance of $\Delta$DOR for the constant term it is obvious that
this method will not be capable of a standalone detection of any type of annual
modulation.

Next we consider an anomaly which has a fixed angle $\beta_\odot$ with
the Sun--spacecraft direction. Again the only time-variable source of
acceleration systematics is the uncertainty in the solar radiation
force. We assume $\beta_\odot = 15\,$deg, which relates to the value,
that we will consider as the maximal flight angle for the trajectories
in the following section as $\gamma < 90\,{\rm deg} -\beta$. Limiting the
tracking time to one year again, we find that Doppler tracking can
detect this type of anomaly up to 23\,AU for the exploration
paradigm and up to 22\,AU for the piggyback micro spacecraft.  Again
sequential ranging is capable of detecting the modulation term beyond
50\,AU for both paradigms, which is more than sufficient for the
mission types under consideration.

In summary, sequential
ranging proves to be the most powerful tracking technique
for a verification of the Pioneer anomaly. In particular, the
discrimination between the candidate directions of a putative
anomaly can be performed by sequential ranging during the whole
length of the interplanetary trajectories under consideration.

At first sight this result seems in contrast with the common wisdom
that range data are usually inferior in quality to Doppler
data.\cite{Curkendall1} However, the standard situation, in which
precision navigation is most relevant, is that of a planetary
approach. In this case the gravitational field is rapidly changing
along the spacecraft orbit, and ranging data induce larger
navigational errors than Doppler indeed. For the deep-space
situation of the Pioneer-anomaly test, the gravity gradients are
very low, and hence the reliability of sequential ranging data is
much improved.

Doppler data will nevertheless be of high importance for the
measurement.  Only by the comparison of both data types, sequential
ranging {\em and} Doppler,  can one discriminate between a real
acceleration and a blue shift of the radio signal.

$\Delta$DOR showed to be of little use for a test of the Pioneer
anomaly. In particular it cannot resolve the directionality of the
anomaly.  Hence, while it is certainly desirable to have occasional
$\Delta$DOR coverage during the Pioneer anomaly test to verify the
orbit reconstruction of the spacecraft (cf.\ Thornton and
Border\cite{Thornton}), $\Delta$DOR does not play a key role in the
precision determination of the anomaly.

From the analysis of the various tracking techniques, we can also
infer requirements on the trajectory of the spacecraft. A upper limit
on the flight angle is desirable if the anomaly direction is supposed
to be determined. In particular, the lowest order modulation term
signaling a velocity-pointing anomaly is proportional to the cosine of
the flight angle. For flight angles close to $\gamma=90$\,deg, the
ability to distinguish an anomaly along the velocity vector from other
candidate effects is suppressed $\sim r_\oplus/r$. Hence a reduced
flight angle considerably improves the sensitivity to such an effect.

Up to now we have not touched upon an effect which could
crucially degrade our measurement accuracy. The above estimates
assume that the spacecraft remains undisturbed during the
measurement period necessary to detect the anomaly or its
modulation. However, this presupposes that no engine firings are
necessary within the time span of detecting the anomaly, and
furthermore that meteoroid impacts are rare enough to leave us with
enough undisturbed measurement intervals to detect the modulation
signals. Concerning thruster firings this condition is in fact
fulfilled. The major disturbance torque in deep space will be the
solar radiation pressure. Even with a low rotation speed of
0.01\,rpm, which we found beneficial for the exploration missions,
the time span between thruster firings necessary to compensate for
this disturbance will be in the order of months, leaving enough time
to conduct precision measurements of the Pioneer anomaly. For the
Pioneer 10 and 11 missions, no disturbances due to the gravitational
fields of asteroids could be noticed. Hence we can exclude this as a
possible source of disturbance for our measurement. Analysis of the
Pioneer tracking data also demonstrated that noticeable meteoroid
impacts occurred only at a frequency of a few per year. We are not
trying to account for a continuous stream of impacts of small dust
particles that are not visible as single events in the tracking
data. Rather we consider such a stream as a putative source of the
anomaly, which should in turn be recognised from its directionality.

\section{Trajectory design \label{traj}}

We have already discussed how the introduction of a momentum
dependence of the gravitational coupling could explain why the
Pioneer anomaly does not show in the planets ephemerides. Even more
straightforwardly, an amplification of the anomaly at high
velocities could occur if matter on low-eccentricity orbits around
the Sun causes a drag force (note however that there does not seem
to be enough dust available\cite{Anderson:2001sg,Nieto:2005bw}). As a
consequence, it is desirable to conduct the Pioneer anomaly test
along a trajectory having a high radial velocity, i.\,e.\ a
hyperbolic escape trajectory, rather than on a bound orbit.
Otherwise, the choice of the inclination, the argument of perihelion
and the longitude of the ascending node do not affect the test. 

From the data of the Pioneer probes, no precise determination of the
direction of the anomalous force was possible. This mainly followed
from the fact that Doppler tracking is able to determine the velocity
of a spacecraft only in the geocentric direction. In particular, it
was not possible to distinguish between the three major candidate
directions of the anomaly: towards the Sun, towards the Earth, and
along the trajectory. The uncertainty in the on-board generated
accelerations makes it therefore desirable to design the spacecraft
trajectory trying to obtain a large flight angle to facilitates the
distinction between the candidate directions from the analysis of the
tracking data (cf.\ the previous section): unfortunately this
requirement is conflicting with the wish to have high radial velocity
of the spacecraft and fast transfer times. A large flight angle could
be obtained by conducting the Pioneer-anomaly measurement as far
inward in the Solar System as possible. Unfortunately, the last
requirement conflicts with the goal of having the smallest possible
systematics generated by solar radiation pressure. A trade-off between
these conflicting requirements has to be made on a case-by-case basis,
and is here discussed for a number of representative trajectories. As
the Cosmic-Vision Programme of the European Space Agency refers to the
decade 2015-2025, this timespan will be used as a baseline launch date
for the trajectories here considered. Missions to Pluto, Neptune and
Uranus are discussed separately from those to Jupiter and Saturn, as
the distances of the former planets allow for a Pioneer anomaly test
to be conducted by the exploration spacecraft during its long
trip. For the latter two targets one has to resort to using a special
micro spacecraft piggybacked to the exploration spacecraft (cf.\
Sec.~\ref{nonded}.\ref{pig}).

\subsection{Orbiter missions to Pluto, Neptune and Uranus \label{trajpop}}

In this paragraph we discuss the possibility of using putative
exploration missions to Pluto, Neptune and Uranus to perform the
Pioneer anomaly test. We will first consider simple flyby missions to
these outer planets. These kind of missions are not too likely to
happen, as the scientific return of a flyby is quite limited and has
already been exploited in several past interplanetary missions. We
will therefore go one step further and consider orbiter missions
exploiting Nuclear Electric Propulsion (NEP) for a final orbital
capture. The trajectory baseline is that of one sole unpowered
gravity-assist around Jupiter. Many trajectory options and missions
are of course possible for exploring these far planets, see for
example Vasile et al., \cite{vasile} but a single Jupiter swingby is
probably the most plausible baseline in terms of risk and mission
time. The purpose is to show that a Pioneer anomaly test would in
general be possible, on these missions, on the vast majority of the
possible trajectories.  In the considered mission scenario the Pioneer
anomaly test would be performed during the ballistic coast phase after
Jupiter. A good trajectory from the point of view of the Pioneer
anomaly test has the following characteristics (cf.\
Sec.~\ref{PA}.\ref{PAe}):
\begin{enumerate}
\item hyperbolic trajectory,
  \item reduced flight angle $\gamma$ (we will allow at maximum 75 deg)
during the test (allowing easy distinction between the velocity
direction and the spacecraft--Earth direction),
  \item long ballistic phase,
  \item large Sun--spacecraft--Earth angle during the test 
(allowing distinction
between the Earth direction and the Sun direction).
\end{enumerate}
We briefly touch upon the implications of these requirements. From
standard astrodynamics we know that along a Keplerian trajectory we
have the following relation for the flight angle
$$
\cos\gamma={\sqrt p} / \left(r\sqrt{\frac 2r-\frac 1a}\right) ,
$$ where $p$ is the semilatus rectum and $a$ the semi-major axis of
the spacecraft orbit. It is therefore possible to evaluate the flight
angle $\gamma$ at any distance from the Sun by knowing the Keplerian
osculating elements along the trajectory after Jupiter. In particular
we note that highly-energetic orbits (i.e.  fast transfers) lead to
larger values of the angle $\gamma$. This leads to prefer a slower
transfer orbit. However, a low velocity results also in a longer trip
and might cause to a smaller value of the anomaly.  The requirement on
the length of the ballistic arc (an issue for orbiter mission
baselines) also tends to increase the transfer time.  In fact the
on-board propulsion (assumed to be some form of low thrust) could
start to brake the spacecraft much later in a slower trajectory (the
square of the hyperbolic velocity, C3, at arrival on a Lambert arc
gets smaller in these missions for longer transfer times). To have a
large Sun--spacecraft--Earth angle during the test phase implies that
the test has to start as soon as possible after the Jupiter swingby
not allowing for a long thrust phase immediately after the swingby as
would be required by optimising some highly constrained trajectory for
low-thrust orbiter missions. To assess the impact of the
requirements on the trajectory design we conduct a multi-objective
optimisation of an Earth-Jupiter-Planet flyby mission assuming pure
ballistic arcs and an unpowered swingby.  We evaluate the solutions
using the Paretian notion of optimality, that is a solution is
considered as optimal if no other solution is better with respect to
at least one of the objectives. We optimise the C3 at Earth departure
as well as the mission duration (as discussed this parameter is
directly related to the flight path angle and to the ballistic arc
length).  The Earth departure date $t_e$, the Jupiter swingby date
$t_j$, and the Planet arrival date $t_p$ were the decision variables,
the departure date being constrained to be within the Cosmic Vision
launch window, and the arrival date being forced to lie before 2100.

The optimisation was performed using a beta version of
DiGMO\cite{COLORADO} (Distributed Global Multi-objective Optimiser),
a tool being developed within the European Space Agency by the
Advanced Concepts Team. The software is able to perform distributed
multi-objective optimisations with a self-learning allocation
strategy for the client tasks. Differential evolution\cite{READING}
was used as a global optimisation algorithm to build the Pareto
sets. Constraints were placed on the Jupiter swingby altitude
($r_p>600,000$\,km). Planet ephemerides were 
the Jet Propulsion Laboratory Digital Ephemerides 405.

\begin{figure}[ht]
    \centereps{\textwidth}{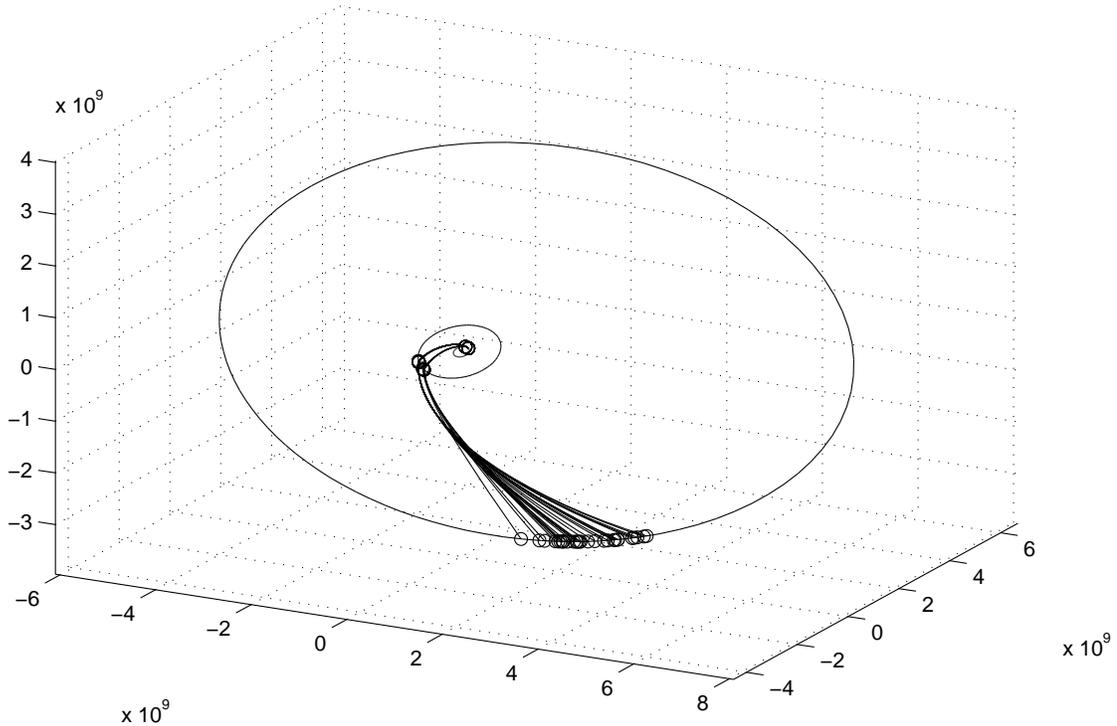}
    \caption{Paretian set for the Earth--Jupiter--Pluto mission within the decade 2015-2025.}
    \label{fig:EJPpareto}
\end{figure}

The results, shown in Fig.~\ref{fig:EJPpareto}, show two main
optimal launch opportunities for the Earth--Jupiter--Pluto transfer:
November 2015 and December 2016. The 2015 launches result in a
slower trajectory (from 17 to 27 years) with lower C3s (of the order
of 87\,${\rm km^2/s^2}$), whereas the 2016 window results in a
shorter mission (from 11 to 15 years) with slightly higher C3s (of
the order of $92$-$100\,{\rm km^2/s^2}$). From a Pioneer-anomaly
test point of view, the only trajectories that would not allow a
good test are the very fast transfers, as the $\gamma$ angle may
become as large as 75~deg by $25$\,AU.  On the other trajectories
the test would be feasible and it would only affect the
low-thrusting strategy, as the test requires a long ballistic arc
with no thrust phase immediately after Jupiter. This requirement is
discussed later.

\begin{figure}[ht]
    \centereps{\textwidth}{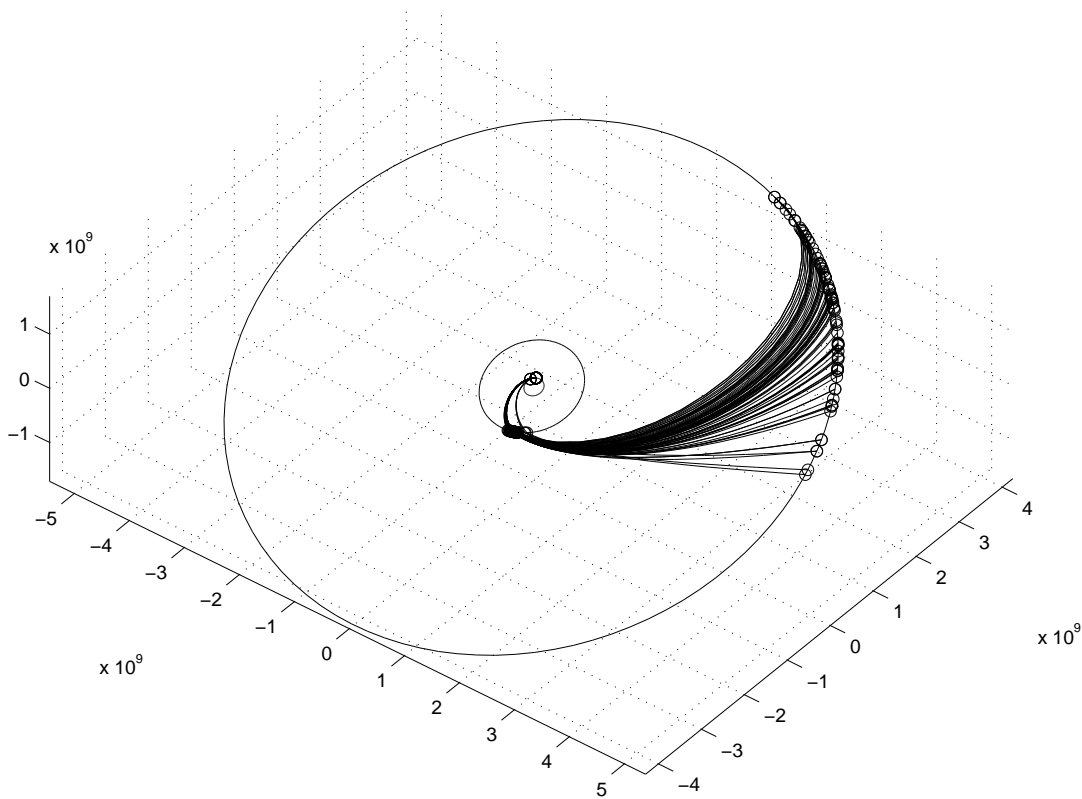}
    \caption{Paretian set for an Earth-Jupiter-Neptune mission within the decade 2015-2025.}
    \label{fig:EJNpareto}
\end{figure}

Similar results are obtained for the Neptune case (see Fig.
\ref{fig:EJNpareto}). There are two optimal launch windows in the
considered decade: January 2018 and February 2019. The first window
allows for very low C3s (of the order of $75\,{\rm km^2/s^2}$) and
transfer times ranging from 14 to 40 or more years, whereas the second
launch window is characterised by higher C3 values (ranging from 90 to
95 ${\rm km^2 / s^2}$) and shorter mission times (as low as 10
years). The requirement on the $\beta$ angle is, in this case,
satisfied by all the trajectories of the Pareto front.

\begin{figure}[ht]
    \centereps{\textwidth}{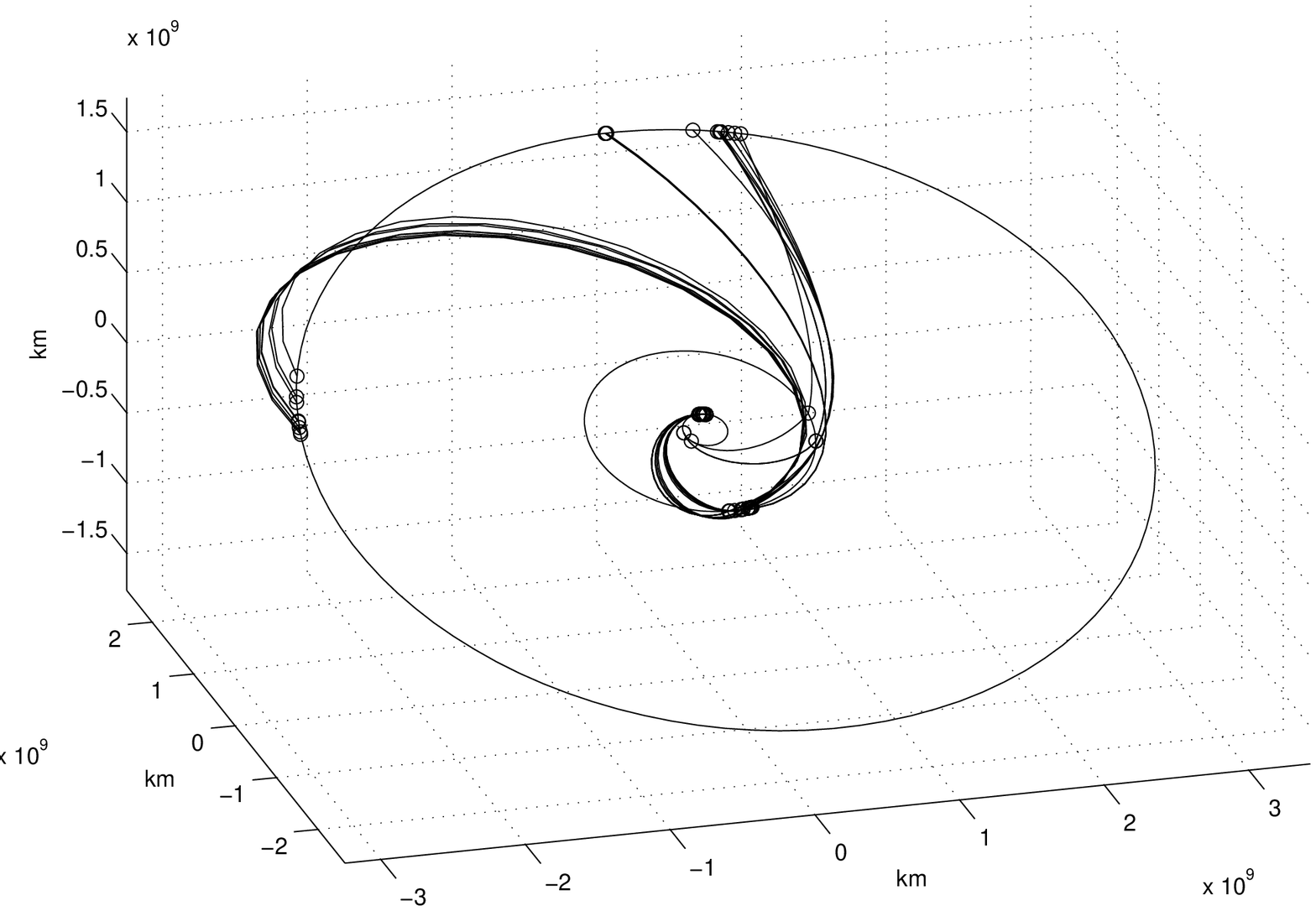}
    \caption{Paretian set for an Earth-Jupiter-Uranus mission within the decade 2015-2025.}
    \label{fig:EJUpareto}
\end{figure}
 The situation for Uranus missions, shown in Fig.
 \ref{fig:EJUpareto}, is slightly more complex. Three main launch
 windows are possible. The first one, corresponding to a late Jupiter
 flyby, is in March 2020 (repeating in April 2021), and corresponds to
 a C3 of roughly $81\,{\rm km^2/s^2}$ (rising to $96\,{\rm km^2/s^2}$
 one year later) and to missions as short as 9 years. The other two
 are in December 2015 and December 2016, producing optimal first-guess
 trajectories with C3s of the order of $78$ to $79\,{\rm km^2/s^2}$
 and transfer times that are either very high (33 years) or of the
 order of slightly more than a decade. Due to the vicinity of the
 planet in this case the value of the flight angle is not an
 issue. Note that a hypothetical mission to Uranus exploiting one
 Jupiter flyby would probably use the 2020 launch opportunity,
 paying an augmented C3 cost of approximately $2\,{\rm km^2/s^2}$ to
 reduce the mission time by several years. The conclusions of the
 preliminary multi-objective optimisation are summarised in
 Table~\ref{launchwindows}.  \renewcommand{\baselinestretch}{1.0}
\begin{table}
\begin{center}
\begin{tabular}{lccc}
 \hline\hline
Target Planet  & Departure Date & Mission Duration& Departure C3\\
& & / years &/ $({\rm km^2/s^2})$\\\hline
Pluto &Nov.~2015& 17-27 & 89-88\\
Pluto &Dec.~2016& 11-15 & 92-100\\
Neptune &Jan.~2018&14-40&74-75\\
Neptune&Feb.~2019&10-12&90-95\\
Uranus&March~2020&9&81\\
Uranus&April~2021&7&96\\
Uranus&Dec.~2015 or Dec.~2016&12-14&79\\
Uranus&Dec.~2015 or Dec.~2016&28-33&79\\
\hline\hline
\end{tabular}
\caption{Pareto-optimal launch windows for flyby missions to the
various outer planets in the considered decade.
\label{launchwindows}}
\end{center}
\end{table}

Each of the trajectories belonging to the Pareto fronts might be
modified to allow an orbiter mission.  Notwithstanding some concepts
to navigate into deep space with solar electric propulsion, it seems
that the nuclear electric option is the most convenient and has to be
used if we want to navigate in the outer regions of the Solar
System. Starting from one of the trajectories of the Pareto-Front, if
the launcher is able to provide all the C3 that is required and we do
not apply heavy constraints, the optimal trajectory will be ballistic
up to the very last phase, and a braking manoeuvre would start just
before the arrival to the planet. If the problem is more constrained,
for example if we add a departure C3 upper limit, then the ion engines
would need to be fired also before and after Jupiter. The firing
immediately after Jupiter is necessary to assure that Pluto orbit is
reached at the right time (this was the case for the POP
trajectory to Pluto\cite{POP1}). In this case a Pioneer anomaly test would
return less scientific data because the thrusting phases could not be
used for the characterisation of the putative anomaly. Adding a
constraint not to use the engines immediately after Jupiter, on the
other hand, would introduce an increase in the propellant mass needed
due to the late trajectory correction. This occurrence would hardly be
accepted by the system designers, and the Pioneer anomaly test would
anyway be possible during the subsequent coast phase of several years.

We may conclude that any trajectory of a flyby or of an orbiter
mission to the outer planets Pluto, Neptune and Uranus is likely to
be suitable for a Pioneer anomaly test with no modifications,
meaning that the three main requirements discussed would be
fulfilled during a trajectory arc long enough to gain significant
insight into the anomaly.

\subsection{Micro spacecraft jettisoned from Jupiter and Saturn missions \label{trajpig}}

\begin{figure}[ht]
    \centereps{\textwidth}{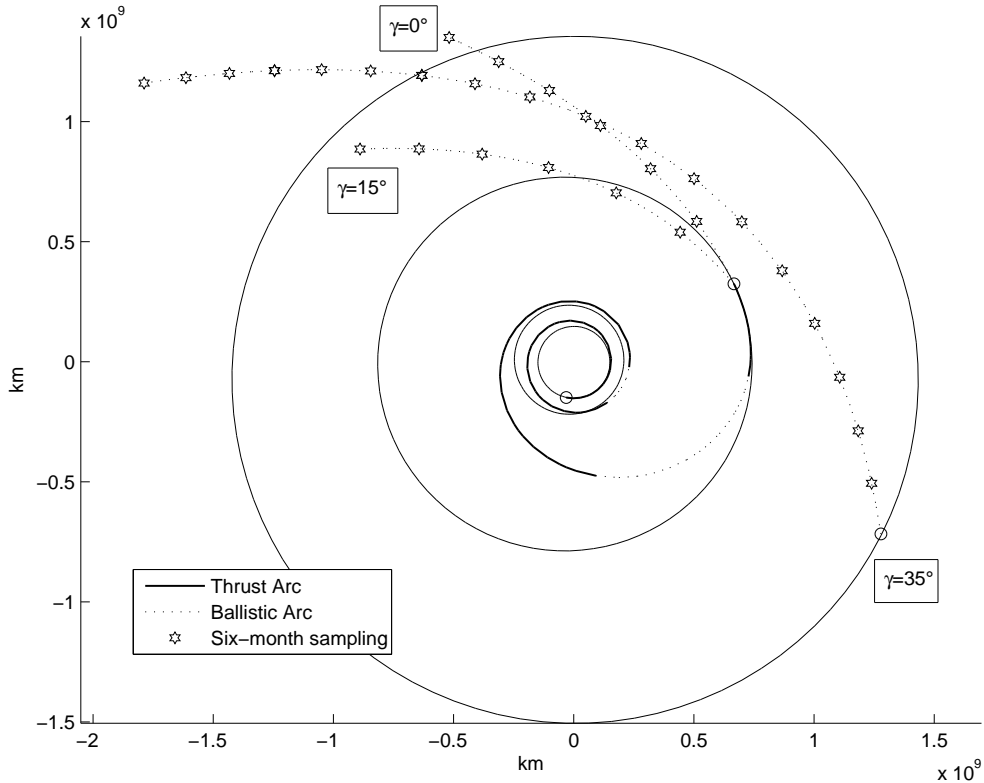}
    \caption{Piggyback micro spacecraft trajectory options.}
    \label{fig:jimoEJ}
\end{figure}

A different situation occurs if we try to test the Pioneer anomaly by
exploiting a putative mission to Jupiter or Saturn. In these cases the
proximity of the planets to the Sun and the likely low energy of the
transfer orbit would not allow for the test to be performed during the
travel to the planet. A possible solution is that of designing a
piggyback micro spacecraft to be added as a payload to the main
mission. We already presented a preliminary assessment of the dry mass
of such a payload in Sec.~\ref{nonded}.\ref{pig} and we now discuss
what the fuel requirement would be on such a spacecraft. As a
guideline for the mother-spacecraft trajectory, we consider the JIMO
baseline and perform an optimisation of a 2016 launch
opportunity. This was done to obtain information on the switching
structure of the thrust so that possible strategies of jettisoning
could be envisaged. The thrust is considered to be fixed and equal to
$2$\,N for a spacecraft weighing $18000$\,kg.  Final conditions at
Jupiter do not take into account its sphere of influence. The
optimised trajectory (visualised in Fig.~\ref{fig:jimoEJ}) foresees a
June 2016 injection into a zero C3 heliocentric trajectory and a
rendezvous with Jupiter in May 2023. We demand that the micro
spacecraft secondary mission does not affect the mothercraft
trajectory, optimised for the main mission goals. A feasible solution
is a spacecraft detaching from the mother spacecraft at the
border of the arrival-planet's sphere of influence, navigating towards
a powered swingby of the target planet, and putting itself
autonomously into a hyperbola of as high as possible energy. Some
general estimates may then be made. We assume that the piggyback
spacecraft is at the border of Jupiter's sphere of influence with zero
C3. The gravity assist has to allow it to gain enough energy to have,
in the heliocentric frame, a hyperbolic trajectory. We also allow for a
non-zero flight angle $\gamma$ at Jupiter.
 Under the assumption of a tangential
burn at the periapsis (cf.\ Gobetz\cite{Gobetz}) we find for the
required $\Delta V$ the expression
\begin{equation}
\label{eq:DVpowered} \Delta
V=\sqrt{V_P^2(3-2\sqrt{2}\cos\gamma)+2\frac{\mu_P}{r_{pP}}}-\sqrt{2\frac{\mu_P}{r_{pP}}}
\, .
\end{equation}
Once the required $\Delta V$ is obtained from Eq.~(\ref{eq:DVpowered})
it is easy to work out the ratio between the propellant mass and the
spacecraft dry mass using the Tsiolkovsky equation. Assuming the use
of chemical propulsion for the powered gravity-assist
($I_{sp}=260$\,s) and putting a constraint on the gravity assist
altitude of $600,000$\,km in the Jupiter case and $40,000$\,km in the
Saturn case, one finds the $\Delta V$ and fuel-to-dry-mass ratio in
dependence of the flight angle as displayed in Table
\ref{tab:JupSat}. Due to the high pericentre required and due to the
greater velocity of the Planet, the Jupiter case requires a higher
propellant mass.

 \renewcommand{\baselinestretch}{1.0}
\begin{table}
\begin{center}

\begin{tabular}{ccccccccccc}
\hline\hline
          &        $\gamma$ / deg                & 0 & 7.5 & 15 & 22.5 & 30 & 37.5 & 45 & 52.5 & 60
          \\\hline

          & $\Delta V$ / (km/s)            & .7 & .8 & 1.1 & 1.6 & 2.2 & 3 & 3.8 & 4.8 & 5.8 \\
  \raisebox{1.5ex}[-1.5ex]{Jupiter case} & $\Delta M/M_0$ & .32 & .37 & .53 & .84 & 1.3 & 2.2 & 3.5 & 5.6 & 8.9
  \\\hline
          & $\Delta V$ / (km/s)             & .17 & .2 & .27 & .39 & .55 & .76 & 1 & 1.3 & 1.6 \\
  \raisebox{1.5ex}[-1.5ex]{Saturn case}  & $\Delta M/M_0$ & .071 & .081 & .11 & .17 & .24 & .35 & .48 & .65 & .86 \\
\hline\hline
\end{tabular}

\end{center}
\renewcommand{\baselinestretch}{2.0}
\caption{Micro spacecraft thrust requirements} \label{tab:JupSat}
\end{table}
 As a consequence, the same spacecraft designed for a $|\gamma|=15
\,\text{deg}$ Jupiter case is capable, in the Saturn scenario, to go
into a $|\gamma|=35\,\text{deg}$ trajectory.  Figure \ref{fig:jimoEJ}
displays example hyperbolic trajectories that first go to decreasing
heliocentric distances and have good performances with respect to
the Pioneer anomaly test. They allow for long periods in
which the direction of the anomaly could be precisely measured, since
the modulations in the tracking signal due to the motion of the Earth,
which enable the determination of the direction of the anomaly, are
enhanced for low heliocentric distances.

\section{Conclusions \label{conc}}

We have considered two plausible mission architectures for the exploration
of the outer Solar System that may also be used to test the Pioneer
anomaly. Firstly a class of low-mass low-thrust missions to Pluto, Neptune or
Uranus. For this mission type the Pioneer anomaly investigation
can be performed by radio-tracking of the exploration spacecraft.
The other mission paradigm considered is that of a
micro spacecraft piggybacked on a large nuclear-reactor-powered
spacecraft sent to explore Jupiter or Saturn. The small spacecraft
would be jettisoned from the mother-craft on the approach to its
destination, would use the target planet of the mother-craft for a
powered swingby, and subsequently performs the Pioneer anomaly
investigation by radio tracking on a hyperbolic coast arc. Starting
from a review of our knowledge of the effect and the models for its
explanation, we have derived a set of minimal requirements for the
spacecraft design and trajectory.

For both mission paradigms the detection of the anomaly is found to be
possible during the whole measurement phase, which extends over
several years.  On-board systematics would still limit the precision
in the determination of the magnitude of the anomaly to approximately
10\%.  This does not seem much of an improvement compared to the 15\%
error margin of the original determination from Pioneer~10 and 11
tracking data (cf.\ above, Sec.~\ref{trac}.\ref{perf}). However by
suitable system design solutions a nondedicated test would be able to
rule out the last candidate onboard sources of the
anomaly. Furthermore the simple requirement of a minimal flight angle
for the trajectories enables the discrimination between the most
plausible classes of candidate models for the anomaly.  The attainable
acceleration sensitivity of $\sim 8\times 10^{-11}\,{\rm m/s^2}$ will
be insufficient for a precise characterisation of the anomaly. In
particular, a slope of the anomaly would most likely only be
determined to the first order -- if at all. This would hardly be sufficient to
determine unambiguously the physical law that might underlie the
Pioneer anomaly.  Hence the quality of the scientific return of
nondedicated missions cannot compete with a dedicated
mission for which acceleration sensitivities down to $10^{-12}\, {\rm
m/s^2}$ would be attainable.\cite{Dittus:2005re} In view of the
ongoing controversial discussion about the origin of the Pioneer
anomaly and the extraordinary costs of a dedicated deep-space mission
to the outer Solar System it seems however more appropriate to consider
the more modest approach of using a nondedicated mission to verify if
the Pioneer anomaly is indeed an indication of a novel physical
effect.\\

\section*{Acknowledgements}

The authors are grateful to Charles Walker at NASA Launch Services
for providing them the JPL ephemerides MATLAB$^{TM}$ routines. This
work has much benefited from discussions with Denis Defr\`ere,
Tiziana Pipoli, Roger Walker, Diego Olmos Escorial, Michael Martin Nieto and Slava G.\ Turyshev.\\

\bibliographystyle{aiaa}
\bibliography{nonded}

\end{document}